\documentclass[amsmath,amssymb,twocolumn,superscriptaddress]{revtex4}

\usepackage{graphicx}
\usepackage{dcolumn}

\def\be{\begin{equation}}
\def\ee{\end{equation}}
\def\beq{\begin{eqnarray}}
\def\eeq{\end{eqnarray}}

\usepackage{bm}
\usepackage{graphicx}
\usepackage{dcolumn}
\usepackage{amsmath}
\usepackage[utf8]{inputenc}
\usepackage{graphicx, psfrag}
\usepackage{amssymb}
\usepackage[colorlinks=true, citecolor=blue, urlcolor = blue, linkcolor= red, bookmarks=true]{hyperref}
\usepackage{float}
\usepackage{amsmath}
\usepackage{tensor}
\usepackage{amsfonts}
\usepackage{dcolumn}
\usepackage{hyperref}
\usepackage{subfigure}
\usepackage{pgfplots}
\usepackage{epstopdf}
\usepackage{booktabs}
\usepackage{amsmath, amssymb}
\usepackage{multirow}
\usepackage{graphicx}
\usepackage{pdflscape}
\usepackage{multirow}
\usepackage{adjustbox}
\usepackage{graphicx}   

\begin{document}

\title{Structure and Mass-Radius Stability of Charged Compact Objects in Symmetric Teleparallel Euler–Heisenberg Gravity}

\author{Allah Ditta}
\email{mradshahid01@gmail.com}
\affiliation{Department of Mathematics, School of Science, University of Management and Technology,  Lahore, 54000, Pakistan.}
\affiliation{Research Center of Astrophysics and Cosmology, Khazar University, Baku, AZ1096, 41 Mehseti Street, Azerbaijan.}

\author{M. Yousaf}
\email{myousaf.math@gmail.com,m.yousaf@vu.edu.pk}
\affiliation{Department of Mathematics, Virtual University of Pakistan, 54-Lawrence Road, Lahore 54000, Pakistan.}

\author{G. Mustafa}
\email{gmustafa3828@gmail.com}
\affiliation{Department of Physics, Zhejiang Normal University, Jinhua 321004, People's Republic of China}

\author{S. K. Maurya}
\email{sunil@unizwa.edu.om}\affiliation{Department of Mathematical and Physical Sciences, College of Arts and Sciences, University of Nizwa 616, Nizwa, Sultanate of Oman}

\author{Farruh~Atamurotov}
\email{atamurotov@yahoo.com}

\affiliation{Kimyo International University in Tashkent, Shota Rustaveli str. 156, Tashkent 100121, Uzbekistan}
\affiliation{Urgench State University, Kh. Alimdjan str. 14, Urgench 220100, Uzbekistan}

\author{Orhan~Donmez}
\email{orhan.donmez@aum.edu.kw}
\affiliation{College of Engineering and Technology, American University of the Middle East, Egaila 54200, Kuwait}

\author{Sardor Murodov} 
\email{mursardor@ifar.uz} 
\affiliation{New Uzbekistan University, Movarounnahr Str. 1, Tashkent 100007, Uzbekistan}
\affiliation{Institute of Fundamental and Applied Research, National Research University TIIAME, Kori Niyoziy 39, Tashkent 100000, Uzbekistan}

\begin{abstract}
In this work, we develop a new relativistic model for a charged anisotropic compact star in the framework of modified symmetric teleparallel gravity, namely $f(Q)$-Euler-Heisenberg gravity. By employing the MIT bag model equation of state, we establish a relation between the metric potentials, leading to an exact solution of the field equations for an anisotropic fluid configuration coupled with a non-linear electromagnetic source. The interior spacetime is smoothly matched with the exterior geometry calculated from the theoretical setup of $f(Q)$-Euler-Heisenberg gravity using the Darmois-Israel junction conditions, ensuring the continuity of the metric functions and their derivatives at the stellar boundary. The physical viability of the model is examined through regularity, energy, and causality conditions, all of which are satisfied throughout the stellar interior. A detailed examination of the physical features, namely the energy density, the radial pressure, the tangential pressure, the anisotropy, and the associated equation-of-state parameters, has been carried out for several choices of the parameter $\gamma$. The study highlights how the pressure anisotropy, the propagation speeds of sound, and the Tolman-Oppenheimer-Volkoff balance condition are interconnected, showing that the star remains in mechanical equilibrium only when the gravitational, hydrostatic, electric, and anisotropic contributions counterbalance one another appropriately. The dynamical stability of the configuration is further supported by the requirement $\Gamma > \tfrac{4}{3}$ for the adiabatic index, indicating resilience against small radial perturbations. The plots of compactness, surface redshift, and the mass--radius profiles confirm that all physical quantities behave regularly and vary smoothly throughout the stellar interior. We graphically plotted the mass-radius curves for one star $(PSRJ\; 1614-2230)$, and numerically calculated the predicted radius for five star candidates $(PSRJ\;1614-2230,\;PSR\;J074+6620,\;PSR\;J1959+2048,\;PSR\;J2215+5135,\;GW190814)$.The behavior of the $f(Q)$ term together with the Euler-Heisenberg coupling shows that these modifications introduce an additional stiffness in the effective equation of state. As a result, the theory is capable of supporting stars that are both heavier and more compact than those allowed within general relativity. When the model is tested against the well-measured pulsar $PSR\; J1614-2230$, the predicted mass and radius fall within the range of current observational constraints, demonstrating the physical reliability of the framework.

\textbf{Keywords}: Compact stars; MIT bag model; Modified gravity; $f(Q)$ gravity; Euler-Heisenberg gravity.
\end{abstract}

\date{\today}

\maketitle

\section{Introduction}\label{sec:int}

Einstein's general theory of relativity (GR) revolutionized the concept of gravitation by replacing the Newtonian picture with a geometric interpretation. In this theory, he described gravity as the curvature of spacetime. The GR serves as the cornerstone for astrophysical sciences which permits the study of gravitational collapse, cosmic expansion, and the intricate dynamics of compact astrophysical systems. Some recently collected observational data, particularly those involving large-scale structure surveys, CMB radiation, and Type $I_a$ supernovae~\cite{riess1998observational,perlmutter1999measurements,riess2007new,peebles2003cosmological,bamba2011time}, established that our cosmos is expanding in an accelerating manner. It suggests the presence of an exotic component commonly referred to as dark energy. 
The dark energy appears to dominate the present cosmic dynamics as it is responsible for the repulsive cosmic behavior, whereas dark matter influences the formation and evolution of visible large scale structures throughout the Universe. Among the key factors shaping cosmic expansion, Einstein's cosmological constant remains a central theoretical element, despite its elegance, GR alone struggles to fully explain Universe's accelerated expansion without resorting to hypothetical entities such as dark energy~\cite{capozziello2006dark,nojiri2007introduction,nojiri2011unified,bamba2012dark,bamba2014cosmology,olmo2019stellar}. 

In consequence of Einstein's formulation of GR, numerous modified and extended theories of gravitation that go beyond the classical Einstein framework were developed in an effort to overcome certain conceptual and mathematical challenges inherent in GR. These frameworks extend or generalize the Einstein-Hilbert action to incorporate higher order curvature terms, matter geometry couplings, or torsional contributions and non-metricity. Consequently, exploring these theories has become a significant direction in contemporary theoretical physics, with profound implications for both cosmology and high energy astrophysics \cite{nojiri2005modified,harko2011f,capozziello2011extended,clifton2012modified,sharif2016energy,nojiri2017modified}. 
One of the earliest and most notable attempts made by Weyl in $1918$~\cite{weyl1918sitzungsberichte}, who proposed a theoretical framework aiming to unify the fundamental interactions of gravitation and electromagnetism within a single geometric scheme. In $1928$ Einstein \cite{einstein1928riemannian,einstein2918neue} motivated by similar unification ideas and explored an alternative approach based on Weitzenb\"{o}ck geometry, unlike GR, where the spacetime dynamics governed by the metric tensor, this formulation employs tetrad fields as the fundamental dynamical variables. Each tetrad field comprises sixteen independent components, leading Einstein to hypothesize that the six components exceeding the ten of the metric tensor could potentially account for the electromagnetic field degrees of freedom. However, subsequent studies revealed that these additional components are not physical in nature but instead correspond to the local Lorentz symmetry intrinsic to the theory as studied in~\cite{muller1983teleparallelism,capozziello2014affine}. Although the early unification attempts of Weyl and Einstein did not succeed in achieving their intended goals. They played a pivotal role in introducing the concept of gauge symmetry into gravitational physics which marked the beginning of extensive efforts to develop gauge formulations of gravity~\cite{raifeartaigh2020dawning}, while Einstein gravity with Gauss-Bonnet entropic corrections investigated in \cite{cognola2013einstein} and study of one loop Euclidean Einstein Weyl gravitational theory in the de Sitter cosmos conducted in \cite{cognola2013one}. In $1979$ building on this perspective, Hayashi and Shirafuji~\cite{hayashi1979new}, proposed a novel gravitational framework known as the New-GR, which interpreted gravity as a gauge theory corresponding to the group of spacetime translations and the formulation of New-GR incorporated three undetermined coupling parameters whose values expected to be constrained through experimental observations.

Within the framework of Weitzenb\"{o}ck geometry, an alternative formulation of gravity known as the Teleparallel Equivalent of GR developed in which the dynamical description of gravitation diverges fundamentally from that of GR, while both theories yield identical field equations, their corresponding action principles are not the same difference arises from a boundary or total divergence term~\cite{maluf2013teleparallel,wu2012matter,capozziello2020weak,el2016exact}.
The Teleparallel Equivalent of GR framework attributes the origin of gravity to spacetime torsion (T), with the curvature tensor constrained to vanish identically. The GR explains gravitational phenomena through spacetime curvature governed by the metric tensor, assuming a torsion free geometry~\cite{li2011f,bahamonde2015modified,cai2016f}. The actions $\int d^{4}x\sqrt{-g}\,T$ reproduce the dynamical content of GR in flat spacetime and also action $\int d^{4}x\sqrt{-g}\,Q$ for nonmetricity ($Q$) equivalent to GR in flat space. The Prominent among these extensions are $f(R)$ gravity, $f(T)$ teleparallel gravity, and the recently developed $f(Q)$ gravity, constructed within the symmetric teleparallel framework among many others. 
Extending Teleparallel Equivalent of GR by promoting $T$ (employing the torsion scalar $T$ as the Lagrangian density instead of curvature) to a general function $f(T)$ leads to the so-called $f(T)$ gravity~\cite{einstein1928riemannian,bengochea2009dark,hanif2024analysis}. This paradigm is especially driven by its capacity to address cosmological obstacles without incorporating exotic matter components. The $f(T)$ gravity has sparked significant attention for its ability to adequately explain both the early Universe inflationary phase and the late-time accelerated expansion. Especially, this gravity produces second-order field equations, which are more technical in nature than the fourth-order equations characteristic of $f(R)$ theories. A substantial body of research explored multiple aspects of $f(T)$ gravity like: classification of different principal categories of this models emphasizing their relevance to cosmic acceleration and null tetrad formalism effectively accommodates various geometries ~\cite{yang2011new,bejarano2015kerr,linder2010einstein,myrzakulov2011accelerating,tamanini2012good,pawar2024two}. Furthermore, Paliathanasis and his collaborators~\cite{paliathanasis2016cosmological} found accurate cosmological solutions for radiation and dust dominated cosmos, and applied their findings to anisotropic Bianchi~I scenarios, demonstrating that isotropic distribution represented by Laurent series techniques. The latter represents a geometrically distinct formulation in which both curvature and torsion vanish, while nonmetricity acts as the sole geometrical descriptor of gravitation, while the symmetric teleparallel approach introduces the nonmetricity scalar $Q$ as the fundamental gravitational invariant. This introduction allows the gravitational action to be generalized to an arbitrary function $f(Q)$ \cite{nester1999symmetric} and this framework preserves the metric as the primary dynamical variable and ensures covariance under coordinate transformations, providing a mathematically consistent platform to investigate gravitational phenomena without invoking curvature or torsion.
Recent studies revealed that $f(Q)$ gravity can successfully reproduce viable cosmological dynamics, mimic the Lambda-CDM expansion history, and account for late-time acceleration, while Maurya et al.~\cite{maurya2024anisotropic} formulated a stellar model using a linear $f(Q)$ action, ensuring regularity and stability through realistic physical conditions and it also employed to analyze stellar equilibrium configurations in \cite{bajardi2020bouncing}. 

In this regard, a useful approach to further improving our comprehension of gravity is to study numerous theoretical models and find observational signs that can discriminate between different approaches, and astrophysical structures occurring in the strong field regime offer natural cosmic research resources for such studies. Potential discriminating features may arise from highly compact objects, such as neutron stars and black holes, or from distinct polarization modes present in gravitational wave signals~\cite{bogdanos2010massive,capozziello2019weak}. From an astrophysical perspective, compact stellar remnants emerge when massive stars deplete their thermonuclear energy sources, leading to the end of internal fusion processes, whereas the subsequent gravitational contraction can result either in the formation of stable, extremely dense configurations or in total collapse, producing black holes. Among these remnants, neutron stars represent objects whose immense density is counteracted by the pressure arising from neutron degeneracy, effectively resisting further gravitational compression, while another well-known category is that of white dwarfs, whose structural stability is sustained by the degeneracy pressure of electrons opposing the inward gravitational force.
From a theoretical perspective, it is well established that the first exact vacuum solution derived within the framework of GR was the Schwarzschild metric~\cite{schwarzschild1999gravitational}, since then, a wide variety of exact and approximate solutions developed to explore the behavior of compact stellar configurations. 
The quest for physically viable interior solutions that accurately depict compact stars emerged as a significant domain for evaluating GR and its modified extensions. Using the right equation of state (EoS), one can look at the stability and balance of compact stars and come to conclusions about what they are made of. The EoS is also very important for looking at the physical properties of stellar matter, especially through the Tolman-Oppenheimer-Volkoff (TOV) equation, which shows the relativistic condition of hydrostatic equilibrium that comes from the interior Schwarzschild solution.

The numerical outcomes of the TOV equations were very important in finding the upper limit on the mass of neutron star systems. These studies show that the upper limit on stable neutron star structures is about $3.2\,M_{\odot}$~\cite{rhoades1974maximum}. This number comes straight from basic physical principles like the principle of causality, the Zeldovich stiff EoS, as well as Le Chatelier's principle. Before these investigations Chandrasekhar~\cite{chandrasekhar1931highly} derived the critical mass limit for white dwarfs, yielding $M_{\mathrm{Ch}} \approx 1.4\,M_{\odot}$, beyond which these stars undergo gravitational collapse. The extensive theoretical analyses, supported by observational evidence, established that the majority of neutron stars possess masses clustered around this characteristic value of $1.4\,M_{\odot}$~\cite{shapiro1983physics,shapiro2024black}. In such compact systems, the degeneracy pressure of neutrons counterbalances the gravitational pull, thereby maintaining hydrostatic equilibrium, while a neutron star with this mass exhibits an average density of approximately $6 \times 10^{14}\,\mathrm{g/cm^3}$ and radius in the range of $10$-$15\,\mathrm{km}$. Some measurements challenged this conventional paradigm, in which observations of pulsars and gravitational wave detections suggest a broader mass spectrum for neutron stars than previously assumed~\cite{horvath2021neutron}. The investigation of gravitational as well as electromagnetic signals from the binary neutron star combined event GW170817 put a strict and essentially model independent limit on the maximum neutron star mass, which is $M_{\text{max}} \leq 2.17\,M_{\odot}$~\cite{margalit2017constraining}. However, later analyses of the identical event~\cite{shibata2017modeling} demonstrate that neutron matter EoS must be sufficiently stiff in order to enable the occurrence of long lived, massive neutron stars with $M_{\text{max}} > 2\,M_{\odot}$, especially in cases the post merger remnant surpasses an initial overall mass of $2.73\,M_{\odot}$. The absence of a relativistic optical counterpart to GW170817 further limits the maximum mass to lie within $M_{\text{max}} \approx 2.15$-$2.25\,M_{\odot}$, while the comparable mass intervals also investigated in \cite{ruiz2018gw170817}. 
Observationally, the pulsar PSR~J1614-2230 measured to possess a mass of $1.928 \pm 0.017\,M_{\odot}$ based on the Shapiro delay effect~\cite{fonseca2016nanograv}, while the millisecond pulsar MSP~J0740+6620 exhibits an even higher mass of $2.14^{+0.10}_{-0.09}\,M_{\odot}$~\cite{cromartie2020relativistic}.

In the framework of GR, the compactness of a self-gravitating stellar configurations is constrained by the well-known Buchdahl limit~\cite{buchdahl1959general,dadhich2020buchdahl}, which imposes the condition $M/R \leq 4/9$, while this bound signifies that beyond a certain mass to radius ratio, a static and stable compact object cannot exist without collapsing into a black hole. An intriguing question that arises is whether alternative or extended gravitational theories can permit configurations exceeding this classical limit, within modified gravitational frameworks, particularly in the context of $f(T)$ or $f(Q)$ gravities, it becomes possible to explore stellar models whose structural properties differ from those predicted by GR. One of the key motivations for such studies lies in determining how the modification of the gravitational action affects the maximum permissible mass for a given EoS, so these alternative formulations may yield different upper mass limits, offering valuable insights into the behavior of ultra dense stellar matter~\cite{margalit2017constraining,harry2020gw190814}. Such investigations are also of astrophysical relevance, for instance, in interpreting events like PSR J0740+ 6620 and GW190814~\cite{nunes2020weighing}, where the secondary component has been proposed as a candidate for an unusually massive neutron star that challenges conventional GR predictions.
These findings provide compelling evidence that neutron stars may attain significantly higher masses than previously predicted by traditional GR-based models, thus motivating the exploration of alternative gravity frameworks like structure, maximum mass, and stable behavior of compact entities in $f(Q)$, $f(Q,T)$ gravitational theory among many others, to account for such massive compact objects \cite{nashed2024structure}.
In the theoretical modeling of stellar structures, it is often convenient to begin with the simplifying assumption of an isotropic matter distribution, implying that the radial and tangential pressures are identical throughout the configuration, however, this idealization does not accurately represent realistic astrophysical scenarios. In actual stellar interiors, the radial pressure generally differs from the tangential one, giving rise to pressure anisotropy, consequently, such systems are characterized by unequal stress components in different spatial directions. In $1933$ the concept of anisotropic stellar models firstly introduced by Lemaitre~\cite{lemaitre1997physics}, furthermore, it demonstrated that for a compact object to achieve a stable equilibrium state, the radial pressure must vanish at the stellar center and decrease monotonically toward the surface, where it becomes zero~\cite{mak2004can}.

A variety of physical mechanisms must be considered when investigating the emergence of anisotropy in compact stellar configurations, but one of the primary factors is the high density regime, in which nuclear interactions can no longer be described by classical physics and must instead be treated within a relativistic scenarios~\cite{ruderman1972pulsars}. The presence of a solid core or the formation of a type-3A superfluid phase may also induce pressure anisotropy within the stellar interior~\cite{kippenhahn1990stellar}. Slow rotational motion represents another potential source of anisotropy~\cite{astashenok2015magnetic}. The emergence of pressure anisotropy in compact stars examined from various theoretical standpoints, notably, the studies by Dev and Gleiser~\cite{dev2003anisotropic} and subsequently by Gleiser and Dev~\cite{gleiser2004anistropic} explored different physical operators capable of producing such anisotropic behavior within stellar interiors. Furthermore, configurations supported by scalar fields, such as boson stars, naturally display anisotropic pressure components~\cite{schunck2003general}. Other exotic astrophysical models including gravastars and traversable wormholes also fall within the class of anisotropic compact systems~\cite{cattoen2005gravastars,de2020general}. Yousaf et al.~\cite{yousaf2025interpretation} explored conservative anisotropic topologically charged wormholes, incorporating relativistic corrections to analyze their stability and energy conditions, while Dai et al.~\cite{dai2025imprints} investigated the imprints of monopole charge in higher-dimensional, revealing how dimensional extensions and anisotropic background modify the distribution of exotic matter. Furthermore, cylindrically symmetric wormholes in the presence of electromagnetic fields, highlighting the dynamical interplay between charge, geometry, and matter anisotropy studied in~\cite{yousaf2023cylindrical}. Anisotropy's role in identifying the structural features of neutron star systems is examined in detail~\cite{bowers1974anisotropic}, demonstrating its significant impact on the equilibrium mass as well as the associated surface redshift, thus affecting observable stellar aspects.  The incorporation of anisotropy in stellar fluid structures is crucial for obtaining accurate representations of massive astrophysical objects, as it affects physical parameters such as internal pressure distribution, mass-to-radius ratios, along with the overall stability of dense star formation. The comprehension of the implications was aided by a number of analytical together with numerical studies conducted within the framework of standard GR and its expanded theories.  In the Buchdahl scenario, for example, Maurya et al.~\cite{maurya2019anisotropic} provided a detailed investigation of anisotropic compact star formation, showing how anisotropy affects the physical feasibility as well as equilibrium of stellar structures. Similarly, Maurya and a few other co-autors~\cite{maurya2016generalised} created a generalized model for anisotropic dense massive stars, and Bhar et al.~\cite{bhar2016modelling} looked into embedding class one geometries to study the internal stellar configuration. Ivanov~\cite{ivanov2017analytical} provided an analytical treatment of anisotropic configurations, emphasizing the role of pressure anisotropy in ensuring stability while, more recently, Kumar and Bharti~\cite{kumar2022review} presented a detailed review summarizing diverse relativistic models for anisotropic compact stars and their physical implications.
Several earlier studies, such as those by Hossein et al.~\cite{hossein2012anisotropic} and Das et al.~\cite{das2021modeling}, examined anisotropic stellar systems in GR with variable cosmological constants, while in the framework of modified gravity, Almutairi et al.~\cite{almutairi2024impact} explored radial perturbations and stability in higher dimensional anisotropic fluid spheres. The dynamical aspects of anisotropic collapse further investigated by Yousaf~\cite{yousaf2025impact} and Yousaf et al.~\cite{yousaf2025dynamical}, who studied the influence of modified curvature functions and hyperbolic corrections on the evolution of compact fluids in $f(R)$ and $f(R,T)$ theories, respectively, providing valuable insights into the nonlinear dynamics of anisotropic self-gravitating systems. The stability properties of anisotropic configurations explored in~\cite{hillebrandt1976anisotropic}, where it showed that their dynamical behavior closely resembles that of isotropic stellar configurations, while Bhatti et al.~\cite{bhatti2022effects} analyzed impact of non-minimal curvature matter coupling on the dynamical behavior of self-gravitating compact fluid configuration in modified gravity. A wide range of anisotropic stellar models proposed to incorporate pressure anisotropy into the energy-momentum tensor, leading to several exact spherically symmetric interior solutions describing realistic cosmic entities~\cite{krori1984some,singh1995new,harko2000anisotropic,esculpi2007comparative,herrera2009expansion,sharma2013relativistic}. 

A comprehensive discussion on the physical origin and implications of pressure anisotropy in compact objects can be found in~\cite{chan2003gravitational,herrera1997thermal}, while, the investigation of supermassive neutron stars within the framework of alternative theories of gravity attracted considerable attention~\cite{astashenok2014maximal,capozziello2016mass}. In the context of teleparallel and symmetric teleparallel theories, several notable investigations extended the modeling of compact stars beyond the Riemannian paradigm, like Ilijic and Sossich~\cite{ilijic2018compact} examined anisotropic stars in $f(T)$ gravity, establishing physically consistent models compatible with observational data. The effects of dark energy components and modified matter fields analyzed by Saha and Debnath in~\cite{saha2019study}, and Chanda et al.~\cite{chanda2019anisotropic} employed the Finch Skea geometry to describe anisotropic configurations in $f(T)$ gravity. With the advent of symmetric teleparallel gravity, where nonmetricity rather than curvature or torsion defines the gravitational interaction, new directions have emerged for modeling dense astrophysical systems. Adeel et al.~\cite{adeel2023physical} carried out a detailed analysis of anisotropic compact stars within $f(Q)$ gravity. Complementary studies by Rani et al.~\cite{rani2024anisotropic} and Bhar et al.~\cite{bhar2024impact} explored the effects of nonmetricity coupling on the stability and energy conditions of anisotropic configurations, demonstrating that $f(Q)$ gravity supports viable compact star solutions consistent with current astrophysical observations.

In order to figure out the interaction between matter geometry connection, energy constraints, as well as the nontrivial topology of universe, it is essential to investigate compact entities such as wormhole geometrical structures within a backdrop of modified teleparallel as well advanced gravitational theories. For example, Ditta et al.~\cite{ditta2021study} comprehensively studied the traversable wormhole solutions in extended teleparallel gravity with matter interaction, showing how coupling settings affect the breach of conventional energy constraints. Ditta et al.~\cite{ditta2025exploring}, examined embedded some compact structures along with their thermodynamic consequences in the same setting, advanced this approach by applying it to phantom wormhole structure and also demonstrating that modified teleparallel interactions are capable of maintaining non-exotic matter distributions demonstrated in \cite{hussain2022phantom}. While Saleem and his collaborators~\cite{saleem2025new} developed new relativistic compact structures in the Rastall teleparallel approach, highlighting the significance of energy constraints in guaranteeing their traversability, Channuie et al.~\cite{channuie2025traversable} investigated wormhole structures with nonlinear electrodynamic amendments, identifying unique gravitational lensing characteristics and energy constraints behaviors in the backdrop of Einstein Euler Heisenberg gravity. Furthermore, how geometric deformations affect the feasibility of such systems by presenting new wormhole solutions based on noncommutative geometry under the influence of extended teleparallel gravity demonstrated in~\cite{ditta2025novel}.
Mandal et al.~\cite{mandal2022study} evaluated the impact of quintessence fields in the formation of anisotropic dense spheres-like structures, revealing how dark energy components influence pressure anisotropy along with equilibrium. Smitha and co-authors~\cite{smitha2025exploring} investigated anisotropy associated with massive compact stars in $f(Q)$ gravity, revealing significant connections between nonmetricity terms and the stiffness of the EoS. These and other advancements in this field broadened the theoretical landscape of $f(Q)$ gravity. In the same manner, Kaur et al.~\cite{kaur2023anisotropic} created a solution that met the vanishing complexity requirement, connecting the stellar configuration's geometric simplicity and physical regularity. These studies highlight the consistent explanation of observed stellar masses, radii, and stability bounds by anisotropic self-gravitating compact fluid configurations in $f(T)$ and $f(Q)$ frameworks.

The structure of the paper is outlined as follows: In Sec.~\ref{sec:bas}, we outlined the basics of the field equations in $f(Q)$-EEH Gravity. In Sec.~\ref{sec:matc}, we studied the evaluation of outer space time and its matching with the inner spacetime to evaluate the unknown parameters involved in the study.
In Sec.~\ref{sec:phy}, we present the discussion about our results and their related physics. In Sec.~5, we present the graphical evolution of mass-radius relation to predict the radius of a star for fixed mass. Our last sec.~\ref{sec:conclusion} suggests the conclusion summary of our study.

 \section{Geometric preliminaries and action}\label{sec:bas}
General Relativity (GR) is traditionally formulated on a spacetime manifold endowed with the Levi-Civita connection. However, the same manifold can accommodate different choices of affine connection, and each alternative connection gives rise to a distinct but physically meaningful gravitational description, offering its own geometric interpretation \cite{ad1,ad2}. To recover the Levi-Civita connection utilized in GR, one must impose the vanishing of torsion and nonmetricity, leaving curvature $R$ as the only geometric property of spacetime. If these restrictions are relaxed, a broader class of gravitational theories can be developed within non-Riemannian geometry, allowing for the coexistence of non-zero curvature, torsion, and nonmetricity.

By choosing an affine connection in which both curvature and nonmetricity vanish, one arrives at the teleparallel equivalent of General Relativity (TEGR) \cite{ad3}. In contrast, symmetric teleparallel gravity arises when the connection is constructed to have vanishing curvature and torsion but non-vanishing nonmetricity \cite{ad4,ad5,ad6}. This geometric setup leads naturally to the formulation of $f(Q)$ gravity, where the gravitational Lagrangian is generalized to an arbitrary function of the nonmetricity scalar $Q$.

Following Jim\'enez \textit{et al.} \cite{ad7}, the $f(Q)$ theory is a symmetric teleparallel extension of GR in which the nonmetricity scalar $\mathcal{Q}$ determines the gravitational action. In the following, we explore the static, spherically symmetric field equations within the framework of $f(Q)$ gravity. The action corresponding to this theory is expressed as:
\begin{eqnarray}\label{1}
S=\int \frac{1}{2} f(Q)\sqrt{-g}\,d^4x + \int (L_m+\mathcal{L}_{\mathrm{EH}}(F, G))\sqrt{-g}\,d^4x,
\end{eqnarray}
where $g$ denotes the determinant of the metric tensor and $L_m$ represents the matter Lagrangian density, and $F \equiv F_{\eta\varrho}F^{\eta\varrho}$ and $G \equiv F_{\eta\varrho}\tilde{F}^{\eta\varrho}$ are the electromagnetic invariants built from the antisymmetric field tensor $F_{\eta\varrho} = \partial_{\eta}A_{\varrho} - \partial_{\eta}A_{\varrho}$. In a general metric-affine spacetime, the nonmetricity tensor is defined as
\begin{eqnarray}\label{a2}
\mathcal{Q}_{k\eta\varrho} = \nabla_k g_{\eta\varrho} = \partial_k g_{\eta\varrho} - \Gamma^{l}_{\eta k} g_{\varrho l} - \Gamma^{l}_{\varrho k} g_{\eta l},
\end{eqnarray}
where $\Gamma^{k}_{\mu\nu}$ denotes the affine connection and $\nabla_k$ is the covariant derivative. The general affine connection can be decomposed as
\begin{eqnarray}\label{3}
\Gamma^{k}_{\eta\varrho} = K^{k}_{\eta\varrho} + L^{k}_{\eta\varrho},
\end{eqnarray}
where $K^{k}_{\eta\varrho}$ and $L^{k}_{\eta\varrho}$ represent the contortion and deformation tensors, respectively, defined by
\begin{eqnarray}\label{4}
L^{k}_{\eta\varrho} = \tfrac{1}{2}\mathcal{Q}^{k}_{\eta\varrho} - \mathcal{Q}^{k}_{(\eta\varrho)}, \quad
K^{k}_{\eta\varrho} = \tfrac{1}{2}T^{k}_{\eta\varrho} + T^{k}_{(\eta\eta)}.
\end{eqnarray}
The antisymmetric part of the affine connection corresponds to the torsion tensor, $T^{k}_{\eta\varrho} = 2\Gamma^{k}_{[\eta\varrho]}$. For the nonmetricity tensor, the superpotential is expressed as
\begin{eqnarray}\label{5}
P^{k}_{\eta\varrho} = \tfrac{1}{4}\left[-\mathcal{Q}^{k}_{\eta\varrho} + 2\mathcal{Q}^{k}_{(\eta\varrho)} + \mathcal{Q}^k g_{\eta\varrho} - \tilde{\mathcal{Q}}^k g_{\eta\varrho} - \delta^{k}_{(\eta}\mathcal{Q}_{\varrho)}\right].
\end{eqnarray}
From the symmetry properties of the metric, two distinct traces of the nonmetricity tensor can be identified:
\begin{eqnarray}\label{6}
\mathcal{Q}_k \equiv \mathcal{Q}_{k\eta}^{\ \ \eta}, \qquad
\mathcal{Q}^k \equiv \mathcal{Q}_{\eta}^{\ k\eta}.
\end{eqnarray}
The nonmetricity scalar $\mathcal{Q}$ is then constructed as
\begin{eqnarray}\label{7}
\mathcal{Q} = -\mathcal{Q}_{\eta\varrho k} P^{\eta\varrho k}
= -g^{mn}\left(L^{\eta}_{\ \varrho n} L^{\varrho}_{\ m\eta} - L^{\eta}_{\ \varrho n} L^{\varrho}_{\ mn}\right).
\end{eqnarray}
The corresponding field equations for $f(Q)$ gravity take the form of
\begin{eqnarray}\label{8}
&&\frac{2}{\sqrt{-g}} \nabla_k\!\Big[\sqrt{-g} f_{\mathcal{Q}} P^{k}_{\eta\varrho}\Big]
+ \tfrac{1}{2} g_{\eta\varrho} f
+ f_{\mathcal{Q}}\!\Big[P_{\eta kl}\mathcal{Q}_{\varrho}^{kl} \nonumber\\
&&- 2\mathcal{Q}_{kl\varrho} P^{kl}_{\eta}\Big]= -T_{\eta\varrho},
\end{eqnarray}
where $f_{\mathcal{Q}} = \partial f / \partial \mathcal{Q}$ and $T_{\eta\varrho}$ denotes the energy\textendash momentum tensor of the matter content. Varying the action (\ref{1}) with respect to the connection yields
\begin{eqnarray}\label{10}
\nabla_\eta \nabla_{\varrho}\!\left(\sqrt{-g} f_{\mathcal{Q}} P^{k}_{\eta\varrho} + H^{k}_{\eta\varrho}\right) = 0,
\end{eqnarray}
where $H^{k}_{\eta\varrho} = -\tfrac{1}{2} \delta L_m / \delta \Gamma^{k}_{\eta\varrho}$ is the density of the hypermomentum tensor.
By imposing the constraint $\nabla_\eta \nabla_{\varrho} H^{k}_{\eta\varrho} = 0$, Eq. (\ref{10}) simplifies to
\begin{eqnarray}\label{11}
\nabla_\eta \nabla_{\varrho}\!\left(\sqrt{-g} f_{\mathcal{Q}} P^{k}_{\eta\varrho}\right) = 0.
\end{eqnarray}

Considering the affine connection's properties, it is convenient to express it in terms of coordinate functions as
\begin{eqnarray}\label{12}
\Gamma^{k}_{\eta\varrho} = \left(\frac{\partial x^{k}}{\partial \xi^{l}}\right) \partial_\eta \partial_{\varrho} \xi^{l},
\end{eqnarray}
where $\xi^{l} = \xi^{l}(x^{\eta})$.
A coordinate system can be chosen in which $\Gamma^{k}_{\eta\varrho} = 0$; this configuration is known as the \textit{coincident gauge}, where the covariant derivative reduces to the partial derivative. Consequently, the nonmetricity tensor in this gauge simplifies to
\begin{eqnarray}\label{13}
\mathcal{Q}_{k\eta\varrho} = \partial_k g_{\eta\varrho}.
\end{eqnarray}
In this formulation, the metric becomes the fundamental dynamical variable, and unlike in GR, the action loses diffeomorphism invariance, a limitation that $f(Q)$ gravity successfully overcomes. In Eq. (\ref{12}), the affine connection is purely inertial, allowing for a fully covariant description independent of gravitational effects \cite{ad8,ad9}.
For astrophysical applications, we now consider a \textit{static, spherically symmetric} spacetime described by the line element
\begin{eqnarray}\label{14}
ds^2 = -e^{\nu(r)}dt^2 + e^{\lambda(r)}dr^2 + r^2(d\theta^2 + \sin^2\theta\,d\phi^2).
\end{eqnarray}

\begin{figure}[!htbp]
\includegraphics[width=8cm, height=5.5cm]{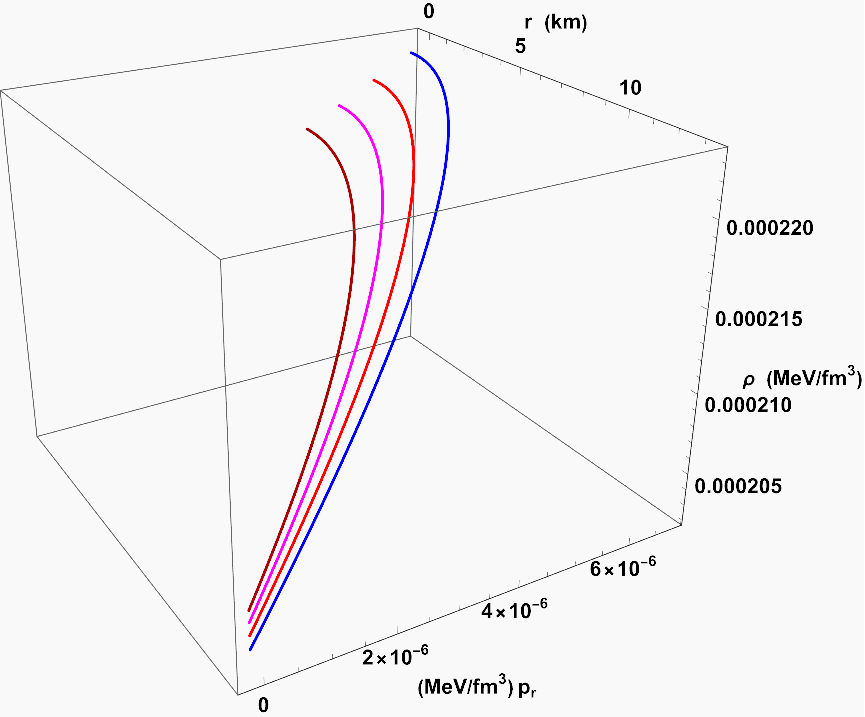}
\includegraphics[width=8cm, height=5.5cm]{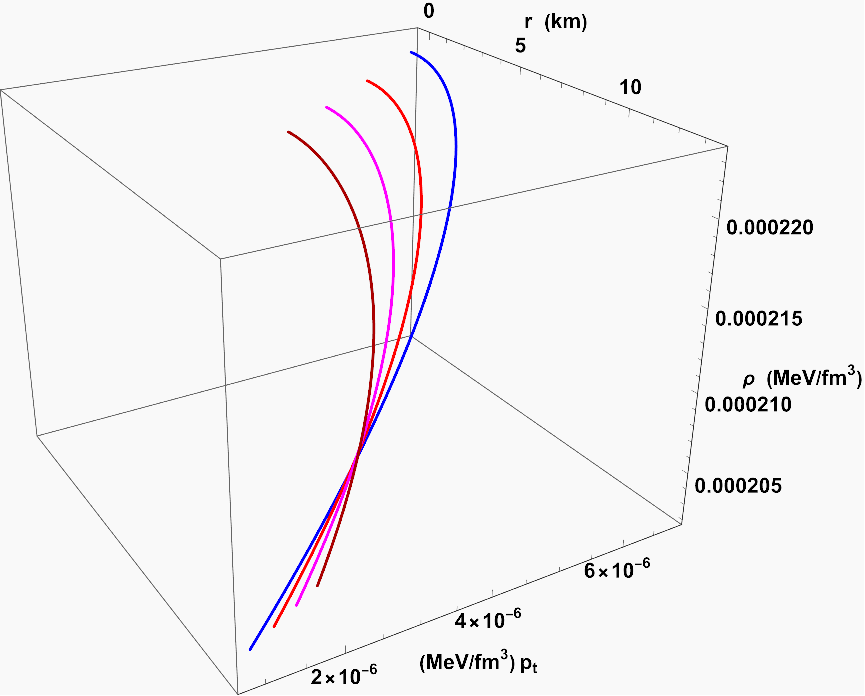}
\includegraphics[width=8cm, height=5.5cm]{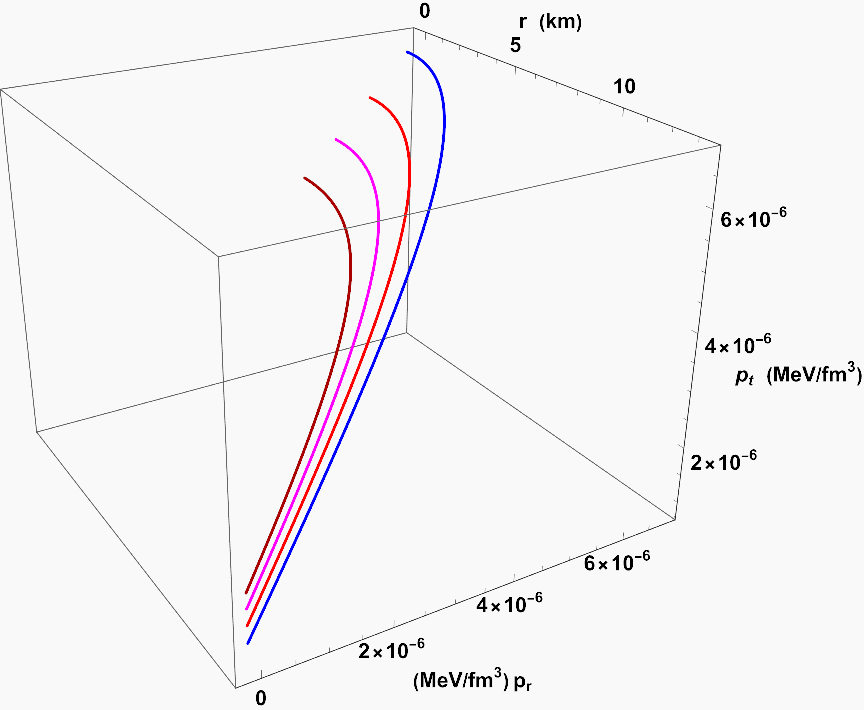}
\caption{ Graphical representation of interdependence of density $\rho$, pressures $p_r$, and $p_t$ along one an other with stellar radius $r$ for fixation of parameters (given in table-\ref{tab1}) and variation of parameters (\textcolor{blue}{$\gamma=-2.3$}, \textcolor{red}{$\gamma=-2.4$}, \textcolor{magenta}{$\gamma=-2.5$}, \textcolor{red!70!black}{$\gamma = -2.6$}) }\label{Fig:1}
\end{figure}

\begin{figure}[!htbp]
\includegraphics[width=8cm, height=5.5cm]{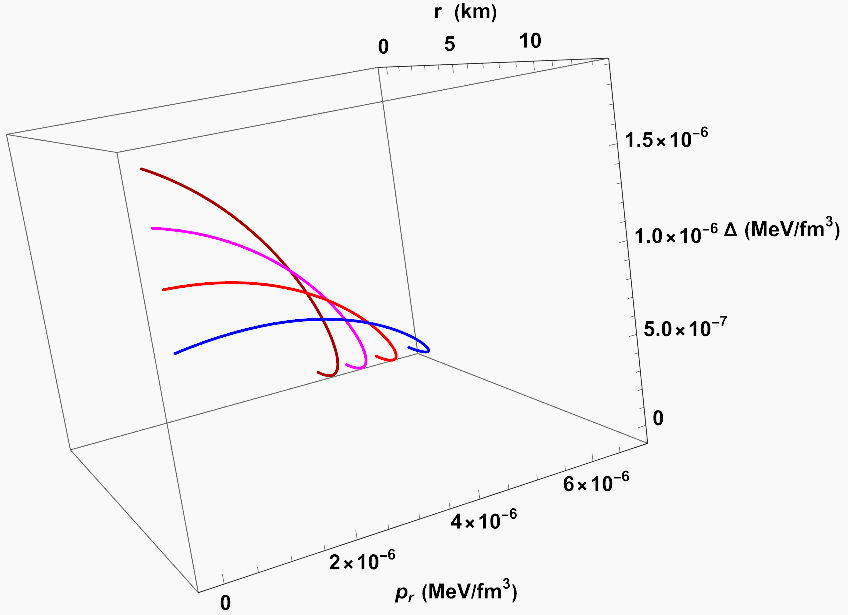}
\includegraphics[width=8cm, height=5.5cm]{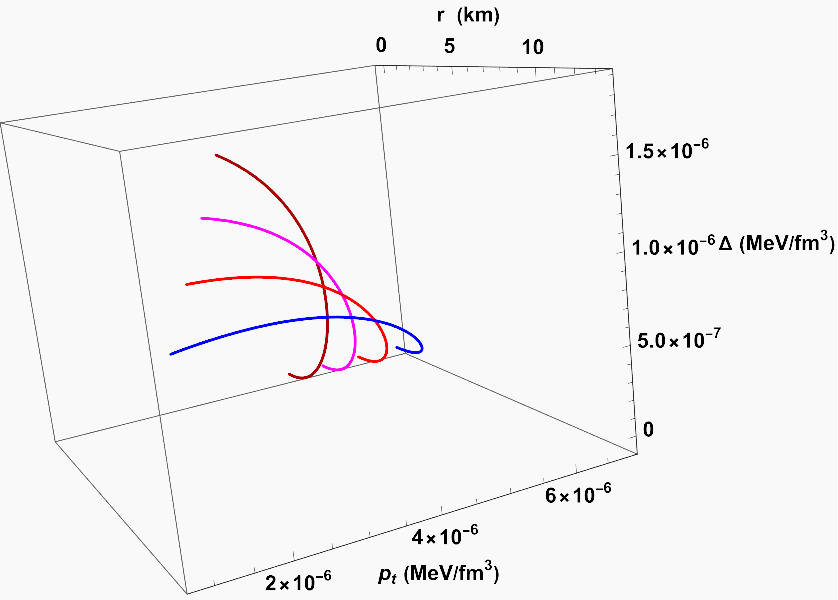}
\includegraphics[width=8cm, height=5.5cm]{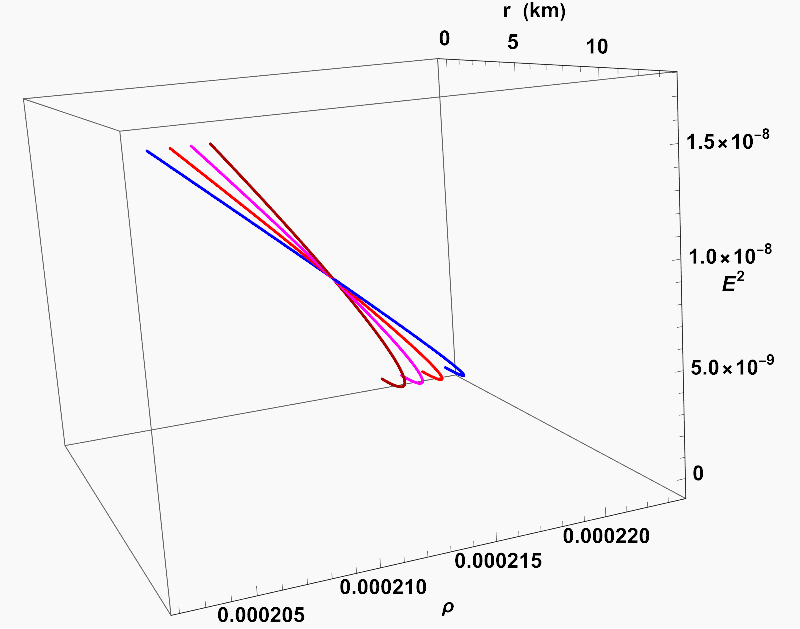}
\caption{Graphical representation of interdependence of anisotropy ($\Delta$) upon density $\rho$, pressures $p_r$, $p_t$, and  radius $r$ for fixation of parameters (given in table-\ref{tab1}) and variation of parameters (\textcolor{blue}{$\gamma=-2.3$}, \textcolor{red}{$\gamma=-2.4$}, \textcolor{magenta}{$\gamma=-2.5$}, \textcolor{red!70!black}{$\gamma = -2.6$})}\label{Fig:2}
\end{figure}

Modeling compact stars in the Einstein-Euler-Heisenberg (EEH) scenario requires solving the $f(Q)$ gravitational field equations in conjunction with the nonlinear electrodynamics sector arising from the Euler-Heisenberg (EH) prescription. This combined treatment enriches the standard formulation $f(Q)$ by embedding quantum electrodynamical corrections, thereby providing a more refined description of strong-field stellar environments. The EH Lagrangian introduces modifications to Maxwell's classical electrodynamics that emerge from vacuum polarization effects occurring under intense electromagnetic fields \cite{heisenberg1936folgerungen,schwinger1951gauge}. The explicit form of the EH Lagrangian density, which accounts for the one-loop quantum corrections, is expressed as \cite{yajima2001black}:

\begin{eqnarray}
\mathcal{L}_{\text{EH}} &=& -\frac{1}{4} F_{\eta\varrho} F^{\eta\varrho} + \alpha (F_{\eta\varrho} F^{\eta\varrho})^2 + \beta (F_{\eta\varrho} \tilde{F}^{\eta\varrho})^2 \\
&=& -\frac{1}{4} F + \alpha F^2 + \beta G^2,
\end{eqnarray}

where the first term represents the classical Maxwellian contribution. The corresponding electromagnetic energy-momentum tensor takes the conventional form:

\begin{equation}
T^{\mathrm{EM}}_{\eta\varrho} = F_{\eta\alpha} F^{\alpha}_{\ \varrho} - \frac{1}{4} g_{\eta\varrho} F_{\gamma\delta} F^{\gamma\delta}.
\end{equation}

The EH Lagrangian $\mathcal{L}_{\mathrm{EH}}$ depends on two electromagnetic field invariants, defined as:

\begin{equation}
F \equiv F_{\eta\varrho} F^{\eta\varrho}, \quad G \equiv F_{\eta\varrho} \tilde{F}^{\eta\varrho},
\end{equation}

where $F_{\eta\varrho}$ and $\tilde{F}_{\eta\varrho}$ denote the electromagnetic field tensor and its dual, respectively. The constants $\alpha$ and $\beta$ encapsulate the QED one-loop correction parameters. Within the EEH formalism, these parameters introduce nonlinear corrections to the electromagnetic sector. The total stress-energy tensor, incorporating both classical and quantum contributions, can be expressed as:

\begin{equation}
T_{\eta\varrho} =  T^{\mathrm{EM}}_{\eta\varrho} + T^{\mathrm{QC}}_{\eta\varrho},
\end{equation}

where $T^{\mathrm{QC}}_{\eta\varrho}$ represents the stress-energy contribution due to quantum effects. The explicit expression for the complete EEH stress-energy tensor reads:

\begin{equation}
T^{\mathrm{QC}}_{\eta\varrho} = g_{\eta\varrho} \mathcal{L}_{\mathrm{EH}} - 4 \frac{\partial \mathcal{L}_{\mathrm{EH}}}{\partial F} F_{\eta\tau} F_{\varrho}^{\ \tau} - 4 \frac{\partial \mathcal{L}_{\mathrm{EH}}}{\partial G} \tilde{F}_{\eta\tau} F_{\varrho}^{\ \tau}.
\end{equation}

The corresponding energy density for the EEH configuration is derived from the temporal component of the above tensor as \cite{channuie2025traversable}:

\begin{eqnarray}
 - T^{0}_{0}{}^{\mathrm{QC}} =\frac{1}{2} (\mathbf{E}^{2} + \mathbf{B}^{2}) + 2 \alpha (\mathbf{E}^{2} - \mathbf{B}^{2})^{2} + 2 \beta ( \mathbf{E} \cdot \mathbf{B} )^{2},
\end{eqnarray}

where $E = F_{tr}$ and $B = \tilde{F}_{tr}$ denote the magnitudes of the electric and magnetic fields. The higher-order terms proportional to $\alpha$ and $\beta$ represent the nonlinear quantum modifications to the classical electromagnetic energy density. The term with $\alpha$ primarily alters the balance between the electric and magnetic field intensities, which can influence the equilibrium conditions by reducing dependence on the matter charge distribution. Similarly, the $\beta$-dependent term contributes only when the electric and magnetic fields are nonorthogonal, i.e., when $\mathbf{E} \cdot \mathbf{B} \neq 0$.

In this setup, $\rho$, $p_t$, and $p_r$ correspond to the energy density, tangential pressure, and radial pressure, respectively. Using Eq.~(\ref{14}), the nonmetricity scalar is given by:

\begin{eqnarray}\label{16}
\mathcal{Q} = -\frac{2 e^{-\lambda(r)}\left(1+r \nu^{\prime}(r)\right)}{r^2},
\end{eqnarray}

where the prime denotes differentiation with respect to $r$. The coupled field equations combining Eq.~(\ref{14}), Eq.~(\ref{8}), and the EEH energy-momentum tensor are:

\begin{widetext}
\begin{eqnarray}
8 \pi  \left(\rho + 24 \alpha  E^4 + \frac{3 E^2}{4} \right) &=& \frac{f}{2} - f_Q \Big[ Q + \frac{1}{r^2} + \frac{e^{-\lambda (r)} (\lambda '(r) + \nu '(r))}{r} \Big], \label{26} \\
8 \pi  \left(p_r - 24 \alpha  E^4 - \frac{3 \alpha  E^2}{4} \right) &=& f_Q \left(Q + \frac{1}{r^2}\right) - \frac{f}{2}, \label{27} \\
8 \pi  \left(p_t + 24 \alpha  E^4 + \frac{3 E^2}{4}\right) &=& f_Q \Big[ \frac{Q}{2} - e^{-\lambda (r)} \Big[\left(\frac{\nu '(r)}{4} + \frac{1}{2 r}\right) (\nu '(r) - \lambda '(r)) + \frac{\nu ''(r)}{2} \Big] \Big] - \frac{f}{2}, \label{28} \\
0 &=& \frac{\cot\theta}{2} Q^{\prime} f_{QQ}. \label{29}
\end{eqnarray}
\end{widetext}

The electromagnetic field tensor is defined as:

\begin{equation}\label{11a}
E_{\epsilon \upsilon} \equiv -\frac{2}{\sqrt{-g}} \frac{\delta (\sqrt{-g} L_{e})}{\delta g^{\epsilon \upsilon}},
\end{equation}

and its stress-energy tensor takes the form:

\begin{equation}\label{eq21}
E_{\epsilon \upsilon} = 2 \left(F_{\epsilon \zeta} F_{\upsilon}^{\ \zeta} - \frac{1}{4} g_{\epsilon \upsilon} F_{\zeta \chi} F^{\zeta \chi}\right),
\end{equation}

with the field tensor components given by:

\begin{equation*}
F_{\epsilon \upsilon} = \mathcal{A}_{\epsilon, \upsilon} - \mathcal{A}_{\upsilon, \epsilon}.
\end{equation*}

The Maxwell identities and field equations follow as:

\begin{eqnarray}
F_{\epsilon \upsilon,\zeta} + F_{\zeta \epsilon,\upsilon} + F_{\upsilon \zeta,\epsilon} = 0, \label{22} \\
(\sqrt{-g} F^{\epsilon \upsilon})_{,\upsilon} = \frac{1}{2} \sqrt{-g} j^{\epsilon}. \label{23}
\end{eqnarray}

Here, $\mathcal{A}_{\epsilon}$ denotes the electromagnetic four-potential, and $j^{\epsilon}$ represents the four-current density. In a static, spherically symmetric configuration, the only nonvanishing component of $j^{\epsilon}$ is $j^{0}$, directed radially. Consequently, the static electric field implies that all components of $F_{\epsilon \upsilon}$ vanish except the radial component $F_{01}$, satisfying $F_{01} = -F_{10}$ due to antisymmetry. The electric field intensity $E(r)$ can be obtained from Eq.~(\ref{eq21}) as:

\begin{equation}\label{24}
E(r) = \frac{1}{2r^2} e^{b(r)+a(r)} \int_{0}^{r} \sigma(r) e^{\lambda(r)} r^{2} dr = \frac{q(r)}{r^{2}},
\end{equation}

where $\sigma$ denotes the charge density and $q(r)$ is the total charge enclosed within radius $r$. The charge function scales with the cube of the radius \cite{goncalves2020electrically}, such that:

\begin{equation}\label{25}
q(r) = k r^3,
\end{equation}

where the constant $k$ characterizes the charge intensity, with $k = 0$ corresponding to an uncharged configuration. This formulation provides a flexible model in which the magnitude of electric charge can be tuned by varying $k$.

\begin{figure}[!htbp]
\includegraphics[width=8cm, height=5.5cm]{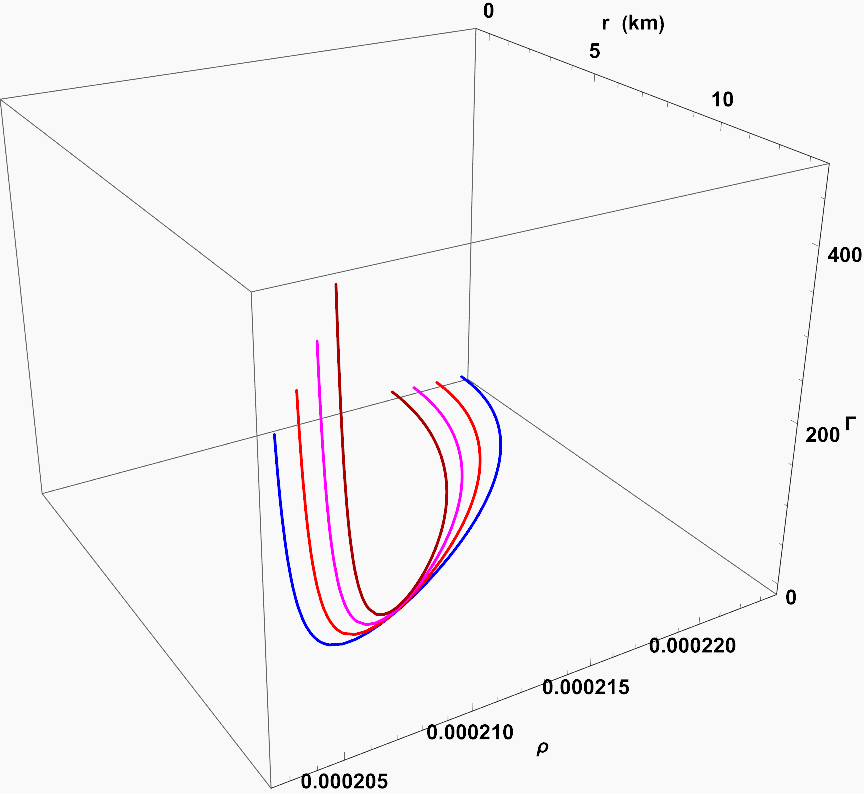}
\includegraphics[width=8cm, height=5.5cm]{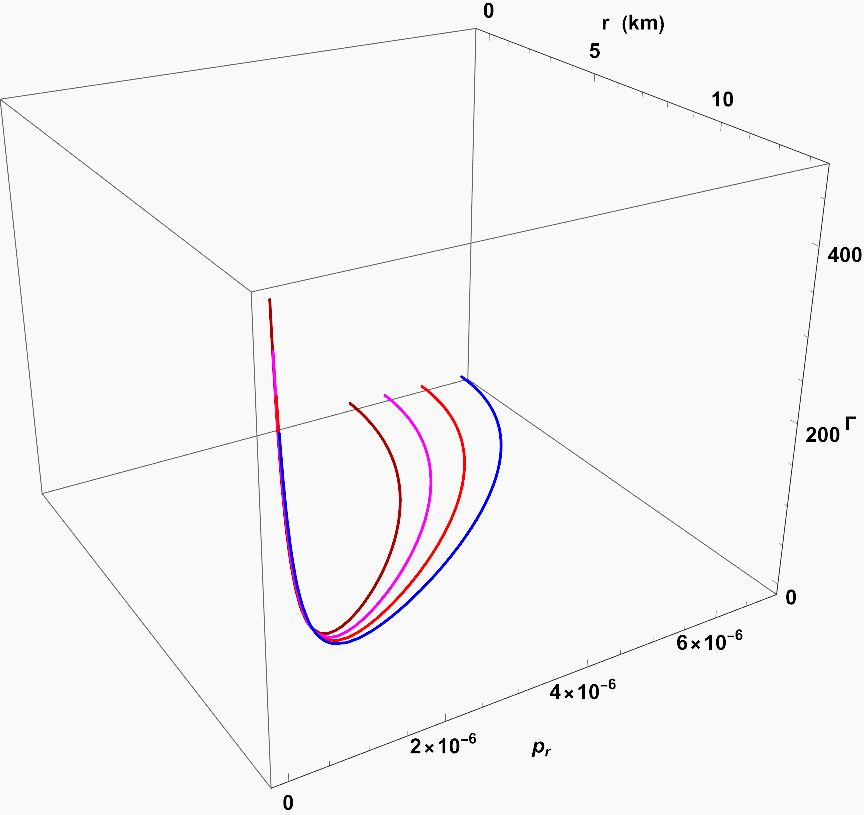}
\caption{Graphical representation of interdependence of Adiabatic index ($\Gamma$) upon density $\rho$, pressures $p_r$, and  radius $r$ for fixation of parameters (given in table-\ref{tab1}) and variation of parameters (\textcolor{blue}{$\gamma=-2.3$}, \textcolor{red}{$\gamma=-2.4$}, \textcolor{magenta}{$\gamma=-2.5$}, \textcolor{red!70!black}{$\gamma = -2.6$})}\label{Fig:3}
\end{figure}

\begin{figure}[!htbp]
\includegraphics[width=8cm, height=5.5cm]{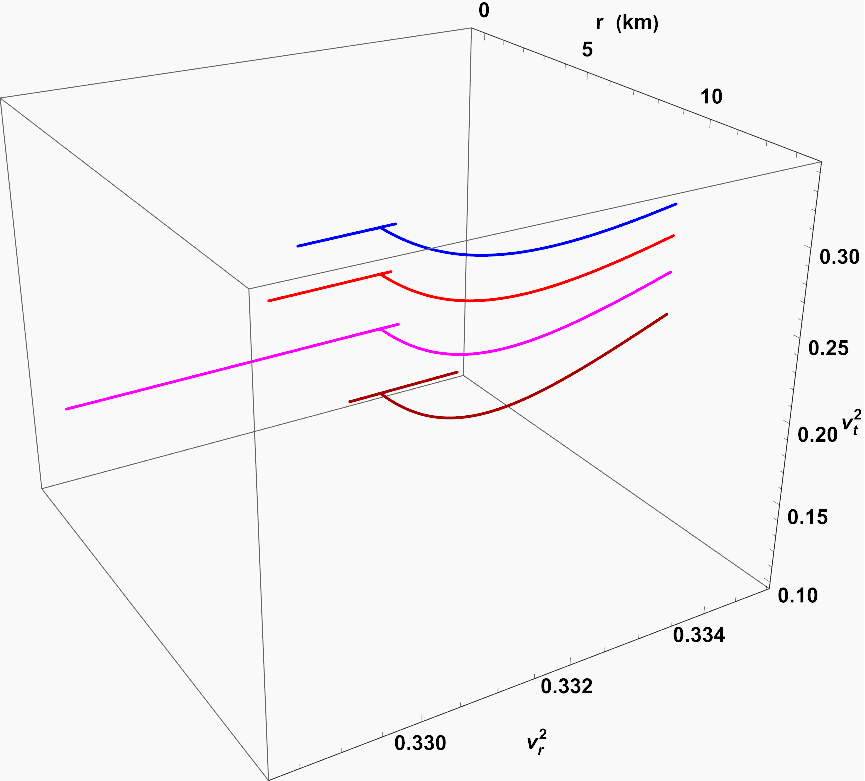}
\includegraphics[width=8cm, height=5.5cm]{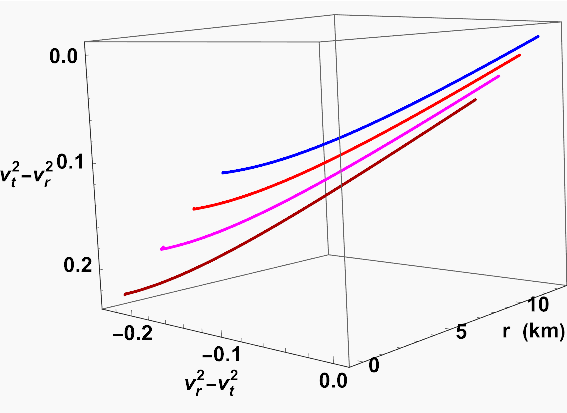}
\includegraphics[width=8cm, height=5.5cm]{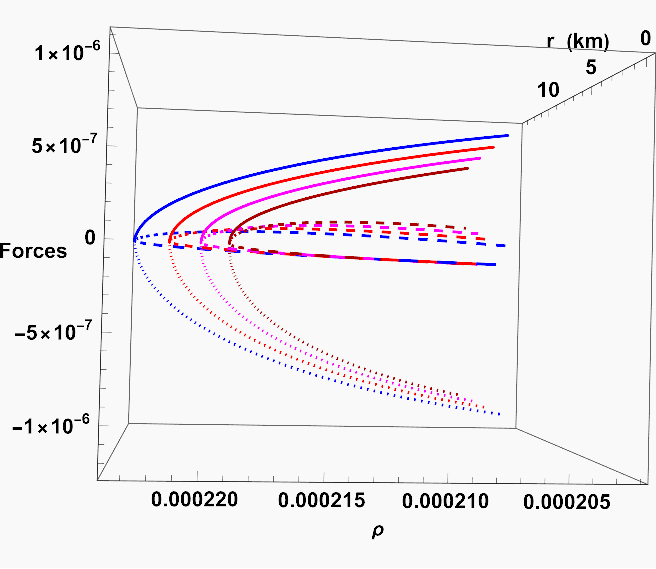}
\caption{Graphical representation of interdependence of sound speeds $v_r^2$ and $v_t^2$ upon each othersand along radial coordinate $r$, and Tov forces upon density $\rho$, and  radius $r$ for fixation of parameters (given in table-\ref{tab1}) and variation of parameters (\textcolor{blue}{$\gamma=-2.3$}, \textcolor{red}{$\gamma=-2.4$}, \textcolor{magenta}{$\gamma=-2.5$}, \textcolor{red!70!black}{$\gamma = -2.6$})}\label{Fig:4}
\end{figure}

\begin{figure}[!htbp]
\includegraphics[width=8cm, height=5.5cm]{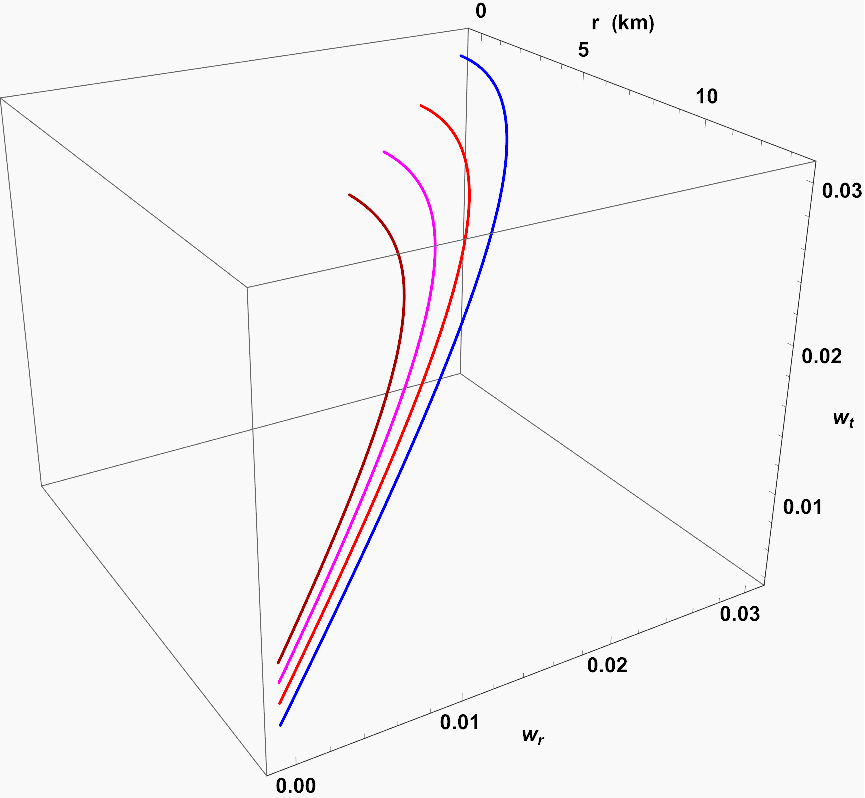}
\includegraphics[width=8cm, height=5.5cm]{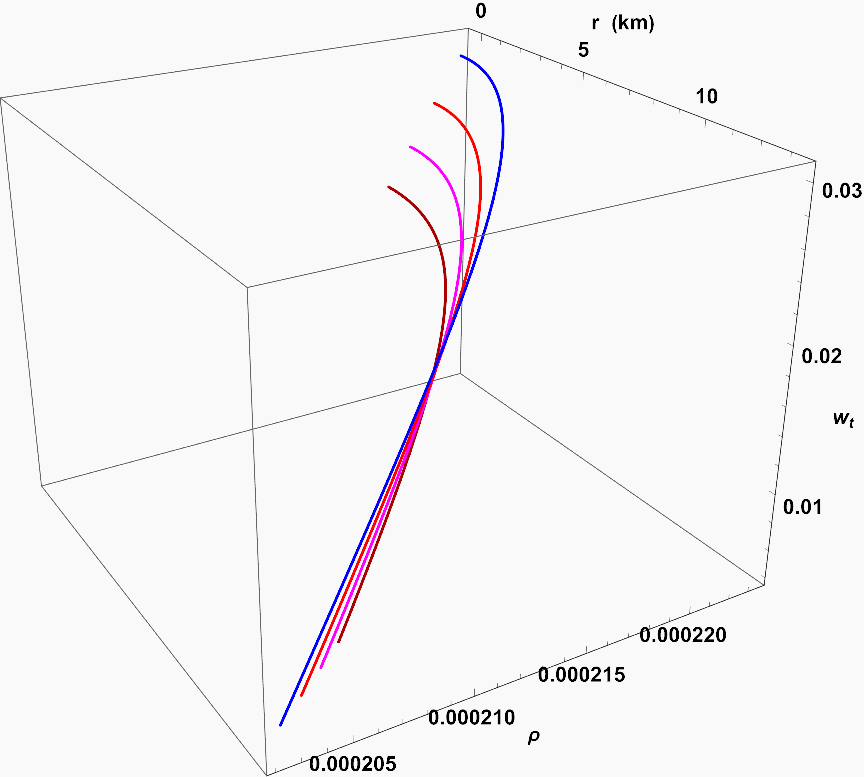}
\includegraphics[width=8cm, height=5.5cm]{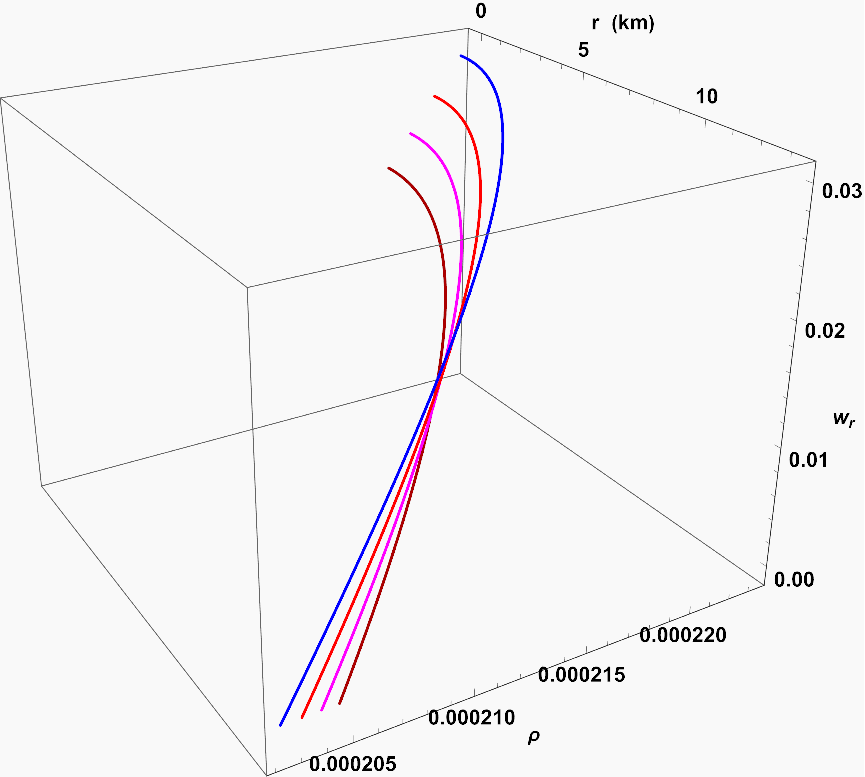}
\caption{Graphical representation of interdependence of equation of state parameters $w_r$ and $w_t$ upon each other, density $\rho$ and radial coordinate $r$ for fixation of parameters (given in table-\ref{tab1}) and variation of parameters (\textcolor{blue}{$\gamma=-2.3$}, \textcolor{red}{$\gamma=-2.4$}, \textcolor{magenta}{$\gamma=-2.5$}, \textcolor{red!70!black}{$\gamma = -2.6$})}\label{Fig:5}
\end{figure}

If the motion equations are combined with Eq.~(\ref{8}), one obtains $f_{QQ} = 0$, which implies that the function $f(Q)$ must be linear. Any attempt to adopt a nonlinear form, such as $f(Q) = Q^2$, results in inconsistencies within the equations of motion and yields nonphysical solutions. In general, nonlinear $f(Q)$ models can only be meaningfully analyzed under a coincident gauge using an extended version of the spherically symmetric metric, as discussed in Ref.~\cite{zhao2022covariant}. In the present work, we adopt a linear form of $f(Q)$ satisfying $f_{QQ}=0$. Integrating Eq.~(\ref{29}) gives the following relation:
\begin{equation}\label{32}
f = \gamma Q + \phi,
\end{equation}
where $\alpha$ and $\phi$ are constants. To determine the physical characteristics of the system, we employ the MIT Bag equation of state (EoS), which provides a realistic representation of quark matter in compact stars. The EoS is given as
\begin{eqnarray}\label{33}
p_r = \omega (\rho - 4 B_g),
\end{eqnarray}
where $B_g$ denotes the bag constant. The MIT Bag model describes quark stars as composed of deconfined quarks confined by a vacuum pressure rather than by gravity alone \cite{li2010bag}. Modern extensions of this model introduce density-dependent bag pressure and vector interactions among quarks, improving the thermodynamic accuracy of strange quark matter (SQM) and up-down quark matter (udQM) \cite{maurya2019study}. These refinements enable better agreement with observed mass-radius relations of compact stars such as EXO~1785-248, PSR~J0740+6620, and GW190814 \cite{ju2024properties, podder2024constraints}. The consistency between the predictions of the MIT Bag model and recent observational data, including gravitational-wave events like GW170817 \cite{joshi2021quark}, strongly supports its applicability in modeling massive quark stars.

In this framework, the field equations (\ref{26})-(\ref{29}) depend on the two metric potentials $\lambda(r)$ and $\nu(r)$. Regularity at the stellar center requires that $e^{\lambda(0)} = 1$, $e^{\nu(0)}$ remains finite, and the first derivatives of both potentials vanish at $r = 0$. These boundary requirements ensure that both the pressure and the energy density attain their highest values at the stellar center and decline smoothly toward the outer surface, a behavior characteristic of a physically stable configuration. With the equation of state~(\ref{33}), the metric function $\nu(r)$ can be expressed as
\begin{eqnarray}\label{ra1} 
\nu(r)&=&\int \Big[\frac{1}{3 \gamma r}\Big[2 e^{\lambda (r)} \Big[r^2 \Big[16 \pi \text{Bg}+3 \pi k^2 r^2 \Big[3 \alpha \nonumber\\&+&128 \alpha k^2 r^2+1\Big]-\phi \Big]+2 \gamma \Big]-4 \gamma \nonumber\\&+&\gamma r \lambda '(r)\Big]\Big] dr
\end{eqnarray}
To fulfill the physical requirements discussed above, we assume a generalized form of the metric potential:
\begin{eqnarray}\label{r2}
e^{\lambda(r)} = 1 + c r^2 e^{a n r^2},
\end{eqnarray}

where $a$, $n$, and $c$ are parameters with dimensions of $\text{length}^{-2}$. Substituting Eq.~(\ref{r2}) into Eq.~(\ref{ra1}) yields
\begin{eqnarray}\label{r1}
\nu (r)&=&A+\frac{1}{6 \gamma }\Big[\frac{2 c e^{a n r^2}}{a^4 n^4} \Big[a^3 n^3 \Big[r^2 \Big[16 \pi \text{Bg}+3 \pi k^2 r^2 \Big[3 \alpha \nonumber\\&+&128 \alpha k^2 r^2+1\Big]-\phi \Big]+2 \gamma \Big]+a^2 n^2 \Big[\phi -2 \pi \Big[8 \text{Bg}\nonumber\\&+&3 k^2 r^2 \left(3 \alpha +192 \alpha k^2 r^2+1\right)\Big]\Big]+6 \pi a k^2 n \Big[3 \alpha \nonumber\\&+&384 \alpha k^2 r^2+1\Big]-2304 \pi \alpha k^4\Big]\nonumber\\&+&2 \gamma \log \left(c r^2 e^{a n r^2}+1\right)+r^2 \Big[32 \pi \text{Bg}+\pi k^2 r^2 \nonumber\\&\times& \left(9 \alpha +256 \alpha k^2 r^2+3\right)-2 \phi \Big]\Big].
\end{eqnarray}
The chosen form of the potential, $e^{\lambda(r)} = 1 + c r^{2} e^{a n r^{2}}$, remains regular and physically acceptable throughout the stellar interior. At the center $r = 0$, the condition $e^{\lambda(0)} = 1$ eliminates any singular behavior in the radial metric component, while the requirement $\left. \frac{d}{dr} e^{\lambda(r)} \right|_{r=0} = 0$ guarantees smoothness at the origin. The temporal component $e^{\nu(r)}$, derived from Eq.~(\ref{r1}), also stays finite and positive for every $r \leq R$, where $R$ denotes the stellar boundary. Its vanishing derivative at the center reflects that both pressure and density attain their peak values at the core and decrease steadily toward the surface. Altogether, these properties confirm a regular geometry and support a stable equilibrium configuration.

To solve the system of Eqs. (\ref{26})-(\ref{28}) for final versions of $\rho$, $p_r$, and $p_t$, using Eqs.(\ref{25}), (\ref{32}), (\ref{r2}), and (\ref{r1}), we obtain the following final expressions:
    \begin{eqnarray}
    \rho &=& \frac{1}{16 r^2 \left(\pi  c r^2 e^{a n r^2}+\pi \right)}\Big[\left(c r^2 e^{a n r^2}+1\right) \Big(r^2 \phi -2 \Big(\gamma \nonumber\\&&+192 \pi  \alpha  k^4 r^6+6 \pi  k^2 r^4\Big)\Big)-\frac{4 c \gamma  r^2 e^{a n r^2} \left(a n r^2+1\right)}{c r^2 e^{a n r^2}+1} \nonumber\\&&+2 \gamma \Big],\label{30}\\
       p_r&=&\frac{1}{48\pi} \bigg(-\frac{2 c \gamma  e^{a n r^2} }{\left(c r^2 e^{a n r^2}+1\right)^2}\left(r^2 \left(c e^{a n r^2}+2 a n\right)+3\right)\nonumber\\&&-384 \pi  \alpha  k^4 r^4-12 \pi  k^2 r^2+\phi -64 \pi\text{Bg}\bigg),\label{31}\\
       p_t&=& -\frac{1}{144 \pi  \gamma  \left(c r^2 e^{a n r^2}+1\right)^3}\Big[512 \pi ^2  \left(c r^2 e^{a n r^2}+1\right)^4 \nonumber\\&\times& \text{Bg}^2 r^2+32 \pi  \text{Bg} \left(c r^2 e^{a n r^2}+1\right)^2 \Big[768 \pi  \alpha  k^4 r^6 \nonumber\\& \times&\left(c r^2 e^{a n r^2}+1\right)^2+6 \pi  (3 \alpha +1) r^4 \left(c k r^2 e^{a n r^2}+k\right)^2\nonumber\\&+&\gamma  \left(c r^2 e^{a n r^2} \left(r^2 \left(4 c e^{a n r^2}+5 a n\right)+15\right)+6\right)\nonumber\\&-&2 \phi  \left(c r^3 e^{a n r^2}+r\right)^2\Big]+2 c^4 r^6 e^{4 a n r^2} \left(r^2 \phi -2 \gamma \right)^2\nonumber\\&-&c^3 r^4 e^{3 a n r^2} \Big(-4 \gamma ^2 \left(5 a n r^2+9\right)+\gamma  r^2 \phi  \nonumber\\& \times&\left(10 a n r^2+37\right)-8 r^4 \phi ^2\Big)+p_{t3}(r)\Big].
   \end{eqnarray}
    where, 
      \begin{eqnarray}
  &&  p_{t1}(r)=\Big[3 \pi  (3 \alpha +1)^2 r^2 \left(c r^2 e^{a n r^2}+1\right)^2+64 \alpha  \Big(\gamma \nonumber\\&& \hspace{1.5cm} \times\left(c r^2 e^{a n r^2} \left(2 r^2 \left(4 c e^{a n r^2}+5 a n\right)+51\right)+33\right)\nonumber\\&& \hspace{1.5cm}-4 \phi  \left(c r^3 e^{a n r^2}+r\right)^2\Big)\Big],\nonumber\\
&& p_{t2}(r)=c \gamma  r^2 e^{a n r^2} \Big(-\left((3 \alpha +1) r^2 \left(4 c e^{a n r^2}+5 a n\right)\right)\nonumber\\&& \hspace{1.0cm}-9 (7 \alpha +3)\Big)+2 (3 \alpha +1) \phi  \left(c r^3 e^{a n r^2}+r\right)^2\nonumber\\&& \hspace{1.0cm}-18 (2 \alpha +1) \gamma ,\nonumber\\
       && p_{t3}(r)=-c^2 r^2 e^{2 a n r^2} \Big(\gamma ^2 \left(2 a n r^2 \left(2 a n r^2-15\right)-30\right)\nonumber\\&&\hspace{1cm}+\gamma  r^2 \phi  \left(20 a n r^2+53\right)-12 r^4 \phi ^2\Big)+294912 \pi ^2 \alpha ^2 k^8 \nonumber\\&&\hspace{1cm}\times r^{10}   \left(c r^2 e^{a n r^2}+1\right)^4+4608 \pi ^2 \alpha  (3 \alpha +1) k^6 r^8 \nonumber\\&&\hspace{1cm}\times \left(c r^2 e^{a n r^2}+1\right)^4+6 \pi  k^4 r^4 \left(c r^2 e^{a n r^2}+1\right)^2 p_{t1}(r)\nonumber\\&&\hspace{1cm}-6 \pi  k^2 r^2 \left(c r^2 e^{a n r^2}+1\right)^2  p_{t2}(r)+c e^{a n r^2} \nonumber\\&&\hspace{1cm}\times \Big(6 \gamma ^2 \left(a n r^2+3\right) \left(2 a n r^2+1\right)-\gamma  r^2 \phi (10 a n r^2\nonumber\\&&\hspace{1cm}+27)+8 r^4 \phi ^2\Big)+\phi  \left(2 r^2 \phi -3 \gamma \right).\nonumber
   \end{eqnarray}

\section{Matching of Interior Metric with the Exterior Metric to Evaluate Unknown Parameters}
\label{sec:matc}

When developing a static and spherically symmetric compact star model, suitable boundary conditions must be imposed to achieve a physically meaningful configuration. These conditions ensure that the interior spacetime connects smoothly with the exterior geometry at the stellar surface. Such a matching not only maintains the continuity of the metric functions but also fixes the integration constants that appear in the interior solution.

The total gravitational mass of the stellar configuration can be obtained by integrating the matter energy density throughout the interior region, which yields
\begin{equation}
M(r) = \int 4 \pi r^{2} \rho \, dr.
\label{eq:m}
\end{equation}
Here, \(M(r)\) denotes the gravitational mass enclosed within a radius \(r\), while \(\rho\) represents the energy density of the matter distribution inside the star.

Using Eq.~\eqref{26} in Eq.~\eqref{eq:m}, the expression for the metric potential corresponding to the exterior spacetime takes the form
\begin{eqnarray}
e^{-\lambda(r)} = 1 + \frac{2M}{\gamma r} + \frac{6\pi k^2 r^4}{5\gamma} + \frac{192\pi \alpha k^4 r^6}{7\gamma} - \frac{r^2 \phi}{6\gamma}.
\label{eq:kk}
\end{eqnarray}

Accordingly, the exterior line element describing the vacuum region surrounding the star can be written as
\begin{eqnarray}
ds_{+}^2 &=& \left(1 + \frac{2M}{\gamma r} + \frac{6\pi k^2 r^4}{5\gamma} + \frac{192\pi \alpha k^4 r^6}{7\gamma} - \frac{r^2 \phi}{6\gamma}\right) dt^2 \nonumber\\
&-& \left(1 + \frac{2M}{\gamma r} + \frac{6\pi k^2 r^4}{5\gamma} + \frac{192\pi \alpha k^4 r^6}{7\gamma} - \frac{r^2 \phi}{6\gamma}\right)^{-1} dr^2 \nonumber\\
&-& r^2 \left(d\theta^2 + \sin^2\theta\, d\phi^2\right).
\label{46}
\end{eqnarray}

At the boundary surface \(r = R\), the interior and exterior spacetimes must join smoothly according to the Darmois-Israel junction conditions, which require the continuity of the metric potentials and the vanishing of the radial pressure at the stellar surface:
\begin{eqnarray}
g_{tt}^{-} = g_{tt}^{+}, \quad g_{rr}^{-} = g_{rr}^{+}, \quad \text{and} \quad p_r(R) = 0.
\end{eqnarray}

These boundary conditions provide a system of equations that can be solved simultaneously to determine the constants appearing in the interior solution. Specifically, the matching ensures that both the gravitational potentials and their physical derivatives are continuous across the surface, guaranteeing the absence of surface layers or discontinuities in the spacetime geometry. The evaluated constants \(A\), \(c\), and \(B_g\) parameters related to the coupling constants (\(\alpha, \gamma, k, \phi\)) can thus be obtained from these matching relations, establishing a complete and physically consistent stellar model. These constants are as:

\begin{widetext}
\begin{eqnarray}
   c&=&-\frac{e^{-a n R^2} \left(5760 \pi  \alpha  k^4 R^7+252 \pi  k^2 R^5+420 M-35 R^3 \phi \right)}{R^2 \left(5760 \pi  \alpha  k^4 R^7+252 \pi  k^2 R^5+420 M-35 R^3 \phi +210 \gamma  R\right)},\\
    B_g&=&\frac{1}{705600 \pi  \gamma  R^4}\Big[20 M \Big[11520 \pi  \alpha  k^4 R^7 \left(a n R^2+1\right)+504 \pi  k^2 R^5 \left(a n R^2+1\right)-35 \Big[2 a n R^5 \phi +R^3 (2 \phi -6 a \gamma  n)\nonumber\\&-&9 \gamma  R\Big]\Big]+R^6 \Big[33177600 \pi ^2 \alpha ^2 k^8 R^8 \left(a n R^2+1\right)+2903040 \pi ^2 \alpha  k^6 R^6 \left(a n R^2+1\right)+1008 \pi  k^4 R^2 \Big[63 \pi \nonumber\\&\times& \left(a n R^4+R^2\right)-400 \alpha  \left(a n R^4 \phi -3 a \gamma  n R^2+6 \gamma +R^2 \phi \right)\Big]-17640 \pi  k^2 \left(a n R^4 \phi -3 a \gamma  n R^2+3 \gamma +R^2 \phi \right)\nonumber\\&+&1225 \phi  \left(-6 a \gamma  n+a n R^2 \phi +\phi \right)\Big]+176400 M^2 \left(a n R^2+1\right)\Big],\\
   A &=&\log \Big[\frac{5760 \pi  \alpha  k^4 R^7+252 \pi  k^2 R^5+420 M-35 R^3 \phi +210 \gamma  R}{210 \gamma  R}\Big]-\frac{1}{6 \gamma }\Big[\frac{2 c e^{a n R^2}}{a^4 n^4} \Big[a^3 n^3 \Big[R^2 \Big[16 \pi  \text{Bg}+3 \pi  k^2 R^2 \nonumber\\&\times&\left(\alpha  \left(128 k^2 R^2+3\right)+1\right)-\phi \Big]+2 \gamma \Big]+a^2 n^2 \left(\phi -2 \pi  \left(8 \text{Bg}+3 k^2 R^2 \left(3 \alpha  \left(64 k^2 R^2+1\right)+1\right)\right)\right)\nonumber\\&+&6 \pi  a k^2 n \left(\alpha  \left(384 k^2 R^2+3\right)+1\right)-2304 \pi  \alpha  k^4\Big]+2 \gamma  \log \left(c R^2 e^{a n R^2}+1\right)\nonumber\\&+&R^2 \left(32 \pi  \text{Bg}+\pi  k^2 R^2 \left(\alpha  \left(256 k^2 R^2+9\right)+3\right)-2 \phi \right)\Big].
\end{eqnarray}    
\end{widetext}
Numerical values of these calculated constants are given in table-\ref{tab1}, for other fixed parameters. 

\begin{table}[h!]
\caption{\label{tab1}Constants evaluated after fixing 
$\alpha =0.01,\;k = 0.00001,\;a=0.001,\;n=0.5,\;\&\;\phi=2.0\times 10^{-24}$ 
by choosing different values of $\gamma$ for following compact star candidates.}
\centering
\begin{tabular}{ccccc}
\hline
\multicolumn{5}{c}{\textit{$PSRJ\;1614-2230$(M = $1.97M/M_{\odot}$, $R= 13.067$ km)\cite{demorest2010two}}} \\
\hline
$\gamma$ & $B_g \; (km^{-2})$ & $c$ & $A$ & $\tfrac{p_{rc}}{\rho_c}$ ($r=0$) \\
\hline
-2.3 & 0.0000506964 & 0.000811354 & -2.49466 & $<1$ \\
-2.4 & 0.0000509047 & 0.000772690 & -2.33290 & $<1$ \\
-2.5 & 0.0000510963 & 0.000737543 & -2.18880 & $<1$ \\
-2.6 & 0.0000512732 & 0.000705455 & -2.05976 & $<1$ \\
\hline
\multicolumn{5}{c}{\textit{$PSR\; J074+6620 $(M = $2.08M/M_{\odot}$, $R= 14.0$ km)\cite{romani2021psr}}} \\
\hline
-2.3 & 0.0000439165 & 0.000686314 & -2.00328 & $<1$ \\
-2.4 & 0.0000440947 & 0.000653676 & -1.87658 & $<1$ \\
-2.5 & 0.0000442588 & 0.000653676 & -1.84402 & $<1$ \\
-2.6 & 0.0000444102 & 0.000596905 & -1.66217 & $<1$ \\
\hline
\multicolumn{5}{c}{\textit{$PSR\; J1959+2048 $(M = $2.18M/M_{\odot}$, $R= 13.98$ km)\cite{kandel2020atmospheric}}} \\
\hline
-2.3 & 0.0000459948 & 0.000727956 & -2.16809 & $<1$ \\
-2.4 & 0.0000461917 & 0.000693094 & -2.02955 & $<1$ \\
-2.5 & 0.0000463728 & 0.000661418 & -1.90604 & $<1$ \\
-2.6 & 0.0000465401 & 0.000632512 & -1.79536 & $<1$ \\
\hline
\multicolumn{5}{c}{\textit{$PSR \;J2215+5135$(M = $2.28M/M_{\odot}$, $R= 13.94$ km)\cite{kandel2020atmospheric}}} \\
\hline
-2.3 & 0.0000482603 & 0.000774284 & -2.35559 & $<1$ \\
-2.4 & 0.0000484781 & 0.000736931 & -2.20346 & $<1$ \\
-2.5 & 0.0000486784 & 0.000703016 & -2.06794 & $<1$ \\
-2.6 & 0.0000488634 & 0.000672085 & -1.94658 & $<1$ \\
\hline
\multicolumn{5}{c}{\textit{$GW190814$(M = $2.67M/M_{\odot}$, $R= 13.58$ km) \cite{abbott2020gw190521} }} \\
\hline
-2.3 & 0.0000596187 & 0.0010198 & -3.42232 & $<1$ \\
-2.4 & 0.0000599488 & 0.00096898& -3.19091 & $<1$ \\
-2.5 & 0.0000602525 & 0.000922986 & -2.98545& $<1$ \\
-2.6 & 0.0000605328 & 0.000881161 & -2.80203& $<1$ \\
\hline
\end{tabular}
\end{table}
\section{Physical Analysis of the Stellar Model}
\label{sec:phy}

Assessing the physical potentials of the model is an important step in determining whether the resulting stellar configuration is compatible with the behavior expected from real compact objects. A principal goal in the study of compact stars is to investigate the internal physical properties that govern their structure and stability. To validate the present model as a realistic stellar configuration, it must satisfy the essential conditions of regularity, causality, and stability outlined earlier. In this section, we examine the physical soundness of the solution obtained in the context of modified $f(Q)$ gravity coupled with Euler--Heisenberg--type nonlinear electrodynamics (the $f(Q)$-EEH framework). The analysis involves checking the behavior of several internal quantities, including the energy density, pressure components, anisotropy, and equilibrium conditions, to ensure that each remains physically acceptable throughout the star’s interior. For numerical comparison, the solution is applied to the well-known star $PSR\;J1614-2230$, whose observed mass and radius provide useful benchmarks for evaluating the viability of the model.

In this work, particular attention is given to understanding how the various stellar characteristics respond within the combined framework of $f(Q)$ gravity and EEH nonlinear electrodynamics. The EEH sector introduces electromagnetic corrections that have a direct impact on the behavior of anisotropy and the distribution of density inside the star. By studying how the geometric modifications introduced through $f(Q)$ interact with the nonlinear electromagnetic field, we aim to clarify the combined influence of these effects on the star’s stability, degree of compactness, and mass-radius relationship. This coupling offers a more complete picture of the internal composition of compact stars and helps illuminate the interplay between modified gravitational dynamics and strong-field astrophysical processes.

The trends displayed in Fig.~\ref{Fig:1} illustrate how the energy density $\rho$, radial pressure $p_r$, and tangential pressure $p_t$ evolve inside the star for various choices of the parameter $\gamma$. All three quantities reach their peak values at the stellar center and then decrease steadily toward the boundary, where the radial pressure drops to zero, allowing for a smooth junction with the exterior spacetime. This monotonic behavior confirms that the matter distribution remains regular and physically acceptable throughout the configuration. The figure further shows that increasing $\gamma$ shifts the curves upward, indicating that larger $\gamma$ values produce higher pressures for the same energy density. This reflects a comparatively stiffer equation of state, which supports more compact and massive stellar objects. The small separation between the $p_r$ and $p_t$ curves reveals a mild degree of anisotropy, with the interior being nearly isotropic near the core and gradually becoming slightly anisotropic as one approaches the outer layers. Overall, the figure highlights how variations in $\gamma$ influence the internal pressure-density structure in the $f(Q)$-EEH framework, reinforcing the regularity and stability of the resulting stellar model.

Figure~\ref{Fig:2} illustrates how the anisotropy factor $\Delta = p_t - p_r$ changes across the interior of the star. At the central point, where $p_r = p_t$, the anisotropy vanishes, indicating a perfectly isotropic core. Since $\Delta$ reflects the difference between tangential and radial stresses, its dependence on $\rho$, $p_r$, and $p_t$ provides useful information about the balance of forces within the star. In the high-density central region, the matter distribution remains nearly uniform, which keeps $p_t$ close to $p_r$ and ensures that $\Delta$ remains close to zero. This guarantees that no directional stress imbalance distorts the central geometry. As one moves outward and the density decreases, the radial pressure drops faster than the tangential component, producing a positive anisotropy ($\Delta > 0$). This outward increasing anisotropy introduces an additional repulsive effect that counteracts gravity, allowing the star to sustain higher masses and greater compactness. A negative anisotropy would do the opposite and weaken the equilibrium. In the present model, $\Delta$ grows smoothly from zero at the center to positive values near the surface, indicating a stable and physically realistic configuration. This behavior reflects the way tangential stresses dominate at lower densities; an expected feature of anisotropic matter in compact stars described by modified gravity models such as $f(Q)$-EEH gravity.

The energy conditions play a central role in examining whether a relativistic stellar model is physically acceptable. As seen from Table~\ref{tab2}, all expressions associated with the different energy conditions decrease smoothly from the center of the star $(r=0)$ toward its boundary $(r=R)$. This decline is expected because both the density and the pressures diminish in the outer layers of a compact object. For each choice of the parameter $\gamma$, the quantities $\rho + p_r$ and $\rho + p_t$, which represent the null and weak energy conditions (NEC and WEC), remain positive throughout the star’s interior. This confirms that the effective energy density exceeds the internal pressures everywhere, ensuring a physically meaningful matter distribution that respects the basic positivity requirements.

The strong energy condition (SEC), expressed through $\rho + p_r + 2p_t$, also retains positive values for all $\gamma$, indicating that the resultant gravitational interaction remains attractive across the configuration. Its gradual decline from the core to the surface mirrors the decrease in density and the corresponding reduction in gravitational pull. The dominant energy condition (DEC), requiring $\rho > |p_r|$ and $\rho > |p_t|$, is similarly satisfied, showing that the energy flow is causal and that no superluminal behavior arises within the interior. The trace energy condition (TEC), given by $\rho - p_r - 2p_t$, remains positive as well, demonstrating that the model does not rely on exotic matter. The gentle drop toward the boundary reflects the increasing role of tangential stresses, consistent with the anisotropic nature of realistic compact stars.

Altogether, Table~\ref{tab2} confirms that all standard energy conditionslike NEC, WEC, SEC, DEC, and TEC are fulfilled everywhere inside the star for every value of $\gamma$ considered. Their monotonic decay toward the outer surface is consistent with a well-behaved physical system in which the density and pressures decrease steadily outward, ensuring stability, causality, and physical admissibility in the setting of $f(Q)$-EEH gravity.

\begin{table}[h!]
\caption{\label{tab2}Values of energy conditions showing monotonic decrease from center to boundary.}
\centering
\begin{tabular}{c|cccc}
\hline
 & expression & center $(r=0)$ & $r=6.5$ & boundary$(r=R)$ \\
\hline
\multirow{5}{*}{\rotatebox{90}{$\gamma=-2.3$}} 
& $\rho+p_r$ & 0.000229406 & 0.00022286 & 0.000202786  \\
& $\rho+p_t$ & 0.000229406 & 0.0002233 & 0.000203767\\
& $\rho-p_r$ & 0.000216096 & 0.000212823 & 0.000202786 \\
& $\rho-p_t$ & 0.000216096 & 0.000212383 & 0.000201804\\
& $\rho-p_r-2p_t$ & 0.000242716 & 0.000233778 & 0.000204749  \\
\hline
\multirow{5}{*}{\rotatebox{90}{$\gamma=-2.4$}} 
& $\rho+p_r$ & 0.000227273 & 0.000221544 & 0.000203619  \\
& $\rho+p_t$ & 0.000227273 & 0.000222057 & 0.000204899\\
& $\rho-p_r$ & 0.000215446 & 0.000212581 & 0.000203619 \\
& $\rho-p_t$ & 0.000215446 & 0.000212068 & 0.000202339\\
& $\rho-p_r-2p_t$ & 0.0002391 & 0.000231533 & 0.000206178  \\
\hline
\multirow{5}{*}{\rotatebox{90}{$\gamma=-2.5$}} 
& $\rho+p_r$ & 0.000225331 & 0.000220338 & 0.000204385  \\
& $\rho+p_t$ & 0.000225331 & 0.00022092 & 0.000205946\\
& $\rho-p_r$ & 0.000214858 & 0.000212362 & 0.000204385 \\
& $\rho-p_t$ & 0.000214858 & 0.00021178 & 0.000202824\\
& $\rho-p_r-2p_t$ & 0.000235803 & 0.000229478 & 0.000207507  \\
\hline
\multirow{5}{*}{\rotatebox{90}{$\gamma=-2.6$}} 
& $\rho+p_r$ & 0.000223555 & 0.000219231 & 0.000205093  \\
& $\rho+p_t$ & 0.000223555 & 0.000219876 & 0.00020692\\
& $\rho-p_r$ & 0.000214324 & 0.000212162 & 0.000205093 \\
& $\rho-p_t$ & 0.000214324 & 0.000211517 & 0.000203266\\
& $\rho-p_r-2p_t$ & 0.000232786 & 0.000227589 & 0.000208746  \\
\hline
\end{tabular}
\end{table}

The adiabatic index (\(\Gamma\)) serves as a crucial stability indicator for compact stellar configurations, as it quantifies the responsiveness of pressure to perturbations in the energy density under adiabatic conditions. In the context of spherically symmetric systems, the analysis of \(\Gamma\) provides valuable insights into the rigidity and dynamical behavior of the stellar matter. For a physically realistic and stable configuration, it is required that the condition \(\Gamma > \frac{4}{3}\) be satisfied throughout the interior of the star \cite{heintzmann1975neutron}. Mathematically, the adiabatic index is defined as  
\begin{equation}
\Gamma = \frac{\rho + p_r}{p_r} \, \frac{d p_r}{d \rho}.
\end{equation}

In the present study, the variation of \(\Gamma\) has been examined for different values of the model parameter \(\gamma\), as shown in figure~\ref{Fig:3}. The figure illustrates the interdependence of the adiabatic index on the energy density \(\rho\), radial pressure \(p_r\), and radial coordinate \(r\). It is observed that \(\Gamma\) takes a smaller value near the stellar core and increases gradually toward the surface. This monotonic growth indicates that the matter distribution becomes progressively stiffer outward, ensuring that the critical stability threshold \(\Gamma > \frac{4}{3}\) is maintained throughout the stellar interior, particularly near the boundary. The observed increase in \(\Gamma\) can be attributed to the interplay between the radial pressure gradient and the decreasing energy density profile. As the energy density decreases outward, the ratio of pressure to density gradually increases, allowing the matter to provide stronger support against the inward pull of gravity. For larger values of the parameter $\gamma$, the adiabatic index assumes slightly higher values, indicating a stiffer equation of state and leading to more compact and stable stellar configurations. The overall behavior of $\Gamma$ illustrated in Fig.~\ref{Fig:3} demonstrates that the system remains dynamically stable under small radial disturbances, confirming that the stellar model developed in the $f(Q)$-EEH framework satisfies the necessary stability requirements expected of realistic compact stars.

In addition, this work also investigates the causality condition within the context of the proposed configuration in $f(Q)$-EEH gravity. This criterion restricts the speeds at which small perturbations propagate through the matter distribution, characterized by the radial and tangential sound speeds, $v_r^{2}$ and $v_t^{2}$. To satisfy the causality requirement, these quantities must obey the inequality

\begin{equation}
0 < v_i^2 < 1, \qquad (i = r, t),
\end{equation}
ensuring that the local signal propagation remains subluminal throughout the stellar interior. The corresponding expressions are defined as
\begin{equation}
v_r^2 = \frac{d p_r}{d\rho}, \qquad v_t^2 = \frac{d p_t}{d\rho}.
\end{equation}

The three-dimensional plots presented in Fig.~\ref{Fig:4} depict how the radial and tangential sound speeds, $v_r^{2}$ and $v_t^{2}$, vary with the radial coordinate for different choices of the parameter $\gamma$. In each case, the velocities stay strictly within the interval $0 < v_i^{2} < 1$. This behavior confirms that the propagation of perturbations remains causal throughout the stellar interior, indicating that the model obeys the fundamental physical requirement on signal speeds.

To further examine the stability of the system, we apply the Herrera--Abreu cracking criterion~\cite{abreu2007sound}, which states that
\begin{equation}
-1 \leq v_t^2 - v_r^2 \leq 1.
\end{equation}
This condition helps determine whether local anisotropies may induce cracking or overturning instabilities in anisotropic matter distributions. As shown in figure~\ref{Fig:4} (middle panel), the difference $v_t^2 - v_r^2$ remains within the prescribed limits for all considered $\gamma$ values, indicating the absence of any cracking instabilities and confirming that the model is locally stable throughout the stellar interior.

In addition, we explore the equilibrium conditions of the compact configuration within the Bardeen geometry under conformal motion, incorporating the Euler--Heisenberg (EH) quantum corrections. The equilibrium of the system is governed by the generalized Tolman--Oppenheimer--Volkoff (TOV) equation, which for an anisotropic and charged fluid in $f(Q)$-EEH gravity takes the form
\begin{eqnarray}
  &&  -\frac{dp_r}{dr} + \frac{2}{r}(p_t - p_r)-\frac{a^{'}(r)}{2}(\rho + p_r)  +\Big[96 \alpha  E^3(r) E^{'}(r)\nonumber\\&&+\frac{3}{2} E(r) E^{'}(r)+\frac{2}{r} \left(16 \alpha  E^4(r)+E^2(r)\right)\Big]=0,
\end{eqnarray}
This can be conveniently expressed as a balance among four distinct force components:
\begin{equation}
\mathcal{F}_a + \mathcal{F}_h + \mathcal{F}_g + \mathcal{F}_{\text{EH}}(r) = 0,
\end{equation}
where
\begin{eqnarray*}
\mathcal{F}_a & =& \frac{2}{r}(p_t - p_r), \;\;
\mathcal{F}_h = -\frac{dp_r}{dr},\\
\mathcal{F}_g &=& -\frac{a^{'}(r)}{2}(\rho + p_r),\\
\mathcal{F}_{\text{EH}}(r) &=&\Big[96 \alpha  E^3(r) E^{'}(r)+\frac{3}{2} E(r) E^{'}(r)\nonumber\\&+&\frac{2}{r} \left(16 \alpha  E^4(r)+E^2(r)\right)\Big].
\end{eqnarray*}

\begin{figure}[!htbp]
\includegraphics[width=8cm, height=5.5cm]{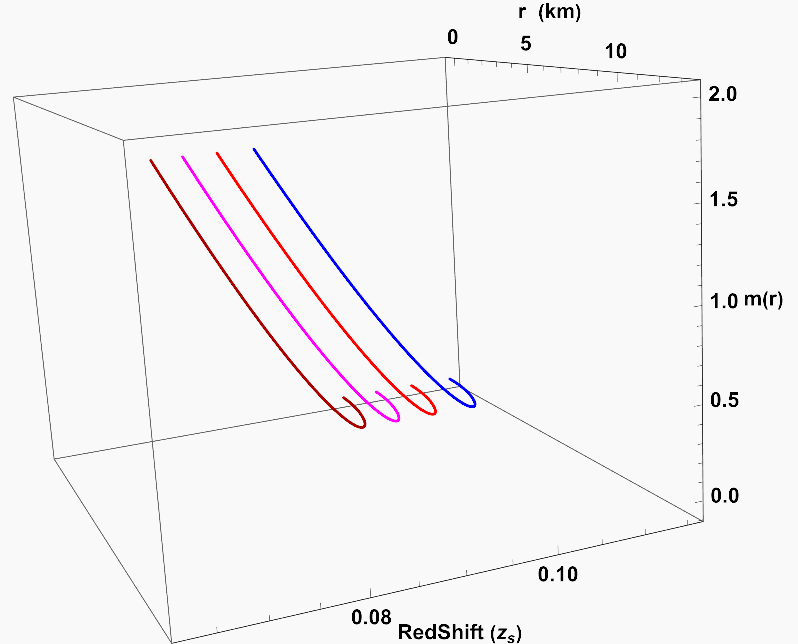}
\includegraphics[width=8cm, height=5.5cm]{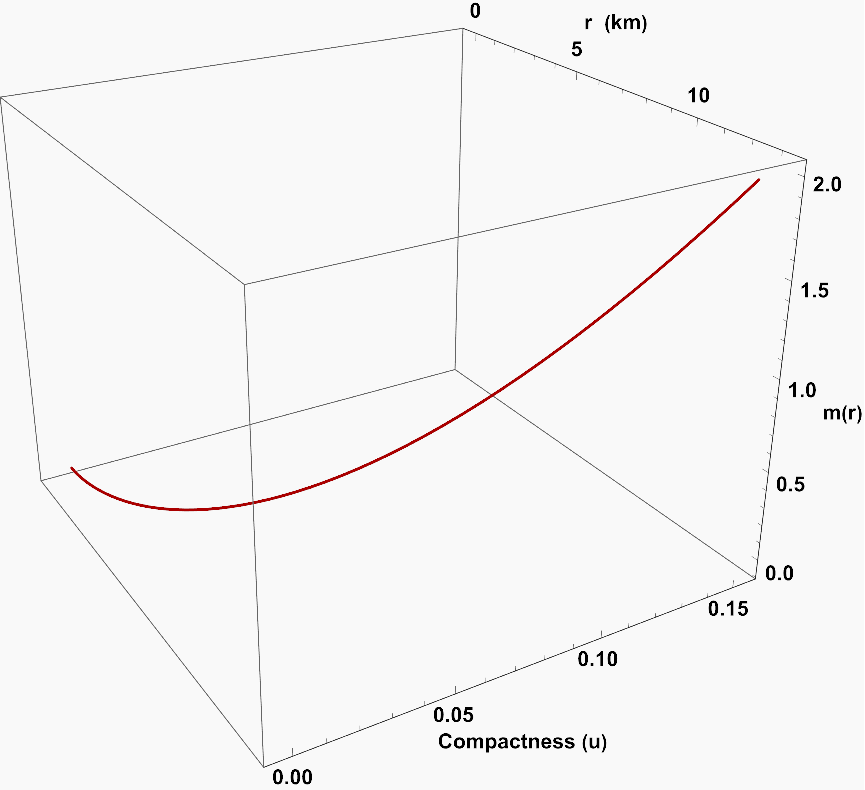}
\includegraphics[width=8cm, height=5.5cm]{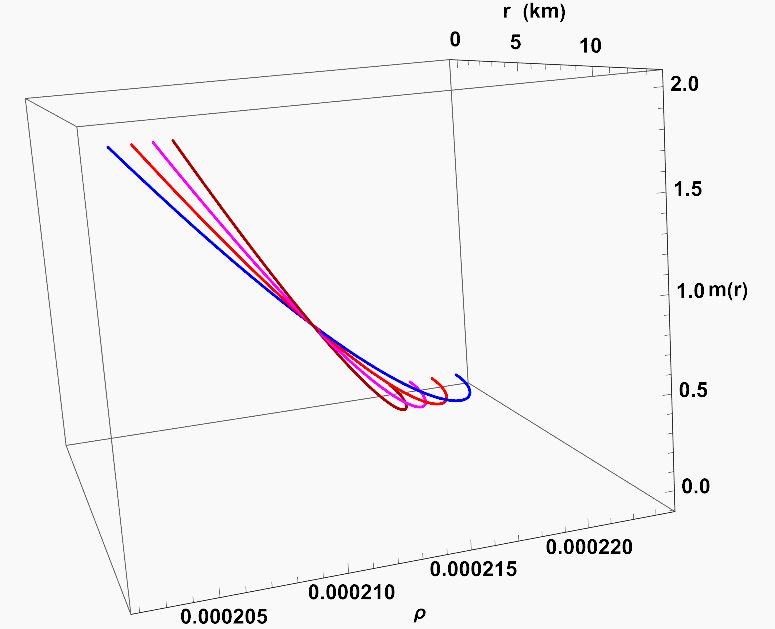}
\caption{Graphical representation of interdependence of mass function on Redshift $z_s$, compactness $u$, and density $\rho$ and radial coordinate $r$ for fixation of parameters (given in table-\ref{tab1}) and variation of parameters (\textcolor{blue}{$\gamma=2.3$}, \textcolor{red}{$\gamma=2.4$}, \textcolor{magenta}{$\gamma=2.5$}, \textcolor{red!70!black}{$\gamma = 2.6$})}\label{Fig:6}
\end{figure}

The three-dimensional visualization in figure~\ref{Fig:4} (bottom panel) reveals the interactive behavior of these forces with respect to the energy density $\rho$ and the radial coordinate $r$. The gravitational and anisotropic forces act inward, while the hydrostatic and EH forces counterbalance them outward. The net effect of all four components remains zero across the stellar radius, demonstrating that the system attains perfect hydrostatic equilibrium. The smooth, symmetric trends in the figure confirm that equilibrium is maintained for all chosen parameter values, ensuring both mechanical stability and physical plausibility of the proposed stellar model under $f(Q)$-EEH gravity.

In order to examine the physical viability of the proposed stellar configuration, we further investigate the equation of state (EoS) parameters that connect the radial and tangential pressures with the energy density of the matter distribution. The EoS parameters are defined as
\begin{equation}\label{58}
w_r = \frac{p_r}{\rho}, \quad w_t = \frac{p_t}{\rho},
\end{equation}
where \( w_r \) and \( w_t \) denote the radial and tangential components of the EoS, respectively. These parameters play a crucial role in understanding the thermodynamic behavior and internal composition of compact objects. For a physically reasonable stellar model, both parameters should satisfy the condition \( 0 < w_r,\, w_t < 1 \), ensuring that the matter distribution corresponds to a normal (non-exotic) fluid and that the pressure remains subdominant to the energy density \cite{das2019new}.

Figure~\ref{Fig:5} presents the three-dimensional variation of \( w_r \) and \( w_t \) with respect to the energy density \( \rho \) and the radial coordinate \( r \) for a fixed set of parameters (as listed in Table~I) and different choices of \( \gamma = 2.3,\, 2.4,\, 2.6 \). The profiles indicate that both \( w_r \) and \( w_t \) remain positive and lie within the expected physical range throughout the stellar interior. The behavior suggests that the matter configuration adheres to a realistic equation of state and excludes the possibility of dark or exotic matter contributions. Moreover, the smooth evolution of \( w_r \) and \( w_t \) with increasing \( r \) implies that the pressure anisotropy is well-behaved and contributes to the overall stability of the system.

An additional feature observed in the profiles of $w_r$ and $w_t$ is that their slopes become progressively steeper as the parameter $\gamma$ increases. This behavior signals a strengthening of the equation of state, meaning that the material inside the star reacts more sharply to changes in density. As a result, configurations corresponding to larger $\gamma$ values develop a more rigid core, enabling them to sustain higher central densities without compromising stability. This sensitivity of the equation-of-state parameters to $\gamma$ highlights the role of microphysical interactions in shaping the internal structure of compact stars and provides useful constraints for selecting physically realistic values of the model parameters.

The compactness factor is another key quantity that characterizes the relativistic behavior and gravitational binding of a compact star. Defined as the ratio of the enclosed gravitational mass to the stellar radius, it measures the influence of the gravitational field on the internal equilibrium and overall structural properties of the configuration. The mass function $m(r)$, which quantifies the amount of mass enclosed within a radial distance $r$, is defined as
\begin{equation}
m(r)=\frac{1}{2}\int \rho(r)\,r^2\,dr.
\end{equation}
Furthermore, the effective mass of the stellar model can be expressed as
\begin{equation}
M_{\textrm{eff}}=\frac{1}{2\gamma}\int_{0}^{R}\left(\rho+\frac{\phi}{2}\right)r^{2}dr=\frac{R}{2}\left[1-e^{-\lambda(R)}\right], \label{eq85}
\end{equation}
which incorporates the geometric modification parameters $(\gamma,\phi)$ appearing in the $f(Q)$-gravity formalism.

Using this effective mass, the compactness factor $u(r)$ is given by
\begin{equation}
u=\frac{M_{\textrm{eff}}}{R},
\end{equation}
and the corresponding surface gravitational redshift is determined as
\begin{equation}
z_s=e^{-\frac{\nu(R)}{2}}-1=(1-2u)^{-1/2}-1.
\end{equation}

Figure~\ref{Fig:6} presents the graphical interdependence of the mass function, compactness, and redshift for different choices of the model parameter $\gamma = 2.3,\, 2.4,\, 2.6$. The plots reveal that the mass function $m(r)$ increases monotonically from the center outward, confirming that the stellar mass accumulates gradually with increasing radius, consistent with the regular behavior of a realistic compact star. The compactness parameter $u(r)$ also shows a smooth and increasing trend, maintaining values within the Buchdahl limit ($u < 4/9$), thereby ensuring that the configuration avoids horizon formation.

The surface redshift $z_s$ follows a similar monotonic behavior, increasing steadily with radius and reaching its largest value at the stellar boundary. This tendency reflects the expected gravitational redshifting of photons escaping from a relativistic compact object. As the parameter $\gamma$ increases, both the compactness $u(r)$ and the corresponding redshift $z_s$ increase in magnitude, indicating that the stellar configuration becomes increasingly relativistic and more tightly bound by gravity. This dependence underscores how sensitive the internal structure of the star is to variations in $\gamma$, with larger values of the parameter leading to denser, more gravitationally intense configurations.

Overall, the observed trends of $m(r)$, $u(r)$, and $z_s$ confirm the physical viability of the proposed model in $f(Q)$ gravity, indicating that the compact star satisfies all necessary regularity and stability conditions throughout its interior.

\begin{widetext}

\begin{table}[h!]
\centering
\caption{Predicted radii of different stellar candidates from the model of $f(T)$-Euler-Heisenberg Gravity for $\rho_s \simeq 2.13\times 10^{14}\,\text{g/cm}^3$.}
\label{table3}

\begin{tabular}{lccccc}
\hline
\multirow{2}{*}{Stellar object} &
\multirow{2}{*}{$M/M_\odot$} &
\multicolumn{4}{c}{Predicted radii (km)} \\
\cline{3-6}
& & $\gamma=-2.3$ & $\gamma=-2.4$ & $\gamma=-2.5$ & $\gamma=-2.6$ \\
\hline

$PSRJ\,1614-2230$ & $1.97$ & 12.75 & 13.50 & 13.75 & 14.25 \\
\hline

$PSR\,J0740+6620$ & $2.08$ & 12.70 & 13.40 & 13.70 & 14.15 \\
\hline

$PSR\,J1959+2048$ & $2.18$ & 12.50 & 13.20 & 13.60 & 14.00 \\
\hline

$PSR\,J2215+5135$ & $2.28$ & 12.25 & 13.10 & 13.40 & 13.90 \\
\hline

$GW190814$ & $2.67$ & 11.75 & 12.25 & 13.00 & 13.50 \\
\hline
\end{tabular}
\end{table}
\end{widetext}

\begin{figure}[!htbp]
\includegraphics[width=8cm, height=7.5cm]{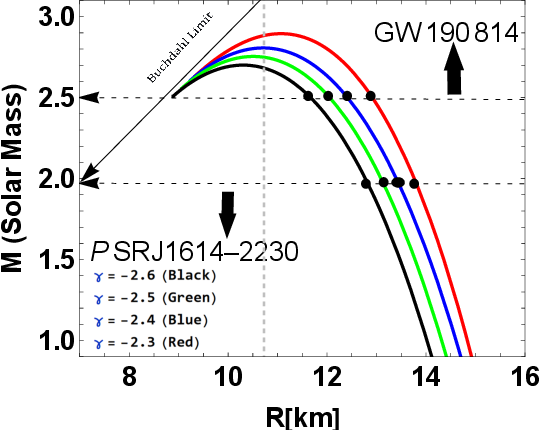}
\caption{Graphical representation of (M-R) curves for different values of $\gamma$. }\label{Fig:mr}
\end{figure}

\section{Mass-Radius curves in $f(T)$-Euler Heisenberg gravity} \label{sec:MR}

 In this section, we examine the mass-radius relation, which plays a central role in assessing the physical viability of our stellar model. The mass-radius curves corresponding to the fixed density $2.13\times 10^{14}\,\text{g/cm}^{3}$ are presented in Fig.~\ref{Fig:mr} for several values of the parameter $\gamma$. For reference, the observationally measured mass of the star $\mathrm{PSR}\, J1614\!-\!2230$ is represented in a graphical analysis, while the radii of four additional compact stars are predicted numerically from the generated mass-radius profiles and summarized in Table~\ref{table3}. The graphs show a clear trend: as $\gamma$ decreases, both the maximum supported mass and the corresponding radius increase, indicating a stiffer effective equation of state within this gravity framework.

\section{Conclusion}\label{sec:conclusion}
In this study, we have developed a comprehensive model of charged anisotropic compact stars within the framework of modified symmetric teleparallel gravity, specifically $f(Q)$-EEH gravity. The metric potentials were related through the phenomenological MIT Bag model equation of state, which effectively characterizes the behavior of strange quark matter at ultra-high densities. The physical quantities such as density, pressure components, and anisotropy are interdependent through the gravitational potentials and electromagnetic field, ensuring that any variation in one parameter dynamically influences the others within the stellar interior. This interplay between the geometric and electromagnetic sectors determines the stability, equilibrium, and degree of compactness of the stellar configuration. To obtain a complete and physically consistent solution, appropriate boundary conditions were imposed so that the interior spacetime could be smoothly joined to the corresponding exterior region. The exterior geometry was derived within the same $f(Q)$-EEH framework using Darmois-Israel matching conditions, ensuring the continuity of the metric components across the stellar surface and allowing the determination of all remaining constants in the model.

A careful analysis of the model shows that all physical variables, like the matter density together with the radial and tangential pressures, remain finite and well behaved throughout the stellar interior. Each of these quantities decreases smoothly from its maximum value at the center toward the boundary, thereby meeting the fundamental requirements of regularity and physical stability. These variables are also closely linked: the gradients of both pressures are determined by the density distribution and the associated gravitational potential, while the electric field and the degree of anisotropy influence the magnitude and slope of the pressure curves. The anisotropy parameter, $\Delta = p_t - p_r$, vanishes in the center and increases outward, reflecting the combined effect of the density profile and the radial coordinate. This outward growth signals a coordinated response between radial and tangential stresses, giving rise to a repulsive anisotropic force that counteracts gravitational attraction and helps maintain hydrostatic equilibrium.

The equation-of-state parameters $w_r = p_r/\rho$ and $w_t = p_t/\rho$ remain within physically reasonable limits $0 < w_r, w_t < 1$, indicating that the distribution of matter adheres to the expected thermodynamic behavior. Their dependence on the density and radial coordinate highlights how pressure adjustments and matter composition evolve coherently across the star. The behavior of the sound speeds $(v_r^2, v_t^2)$ further demonstrates the intricate dynamical connection among these variables. Both sound speeds are within the causal domain $0 < v_r^2, v_t^2 < 1$, and the inequality $|v_t^2 - v_r^2| \leq 1$ is satisfied everywhere, ruling out any local cracking instabilities.

An important aspect of this work is the examination of how the sound speeds relate to the forces present in the generalized Tolman-Oppenheimer-Volkoff equation. The equilibrium study shows that gravitational, hydrostatic, electric, and anisotropic forces counterbalance each other precisely at all interior points. Changing the parameter $\gamma$ modifies this balance, demonstrating how variations in the underlying geometry and coupling with matter influence the distribution of forces, density, and pressure gradients simultaneously.

The adiabatic index, $\Gamma = \frac{\rho + p_r}{p_r}\frac{dp_r}{d\rho}$, remains above the critical value of $\tfrac{4}{3}$ throughout the star, confirming stability against small radial perturbations. Its dependence on $r$ and $\rho$ shows that regions of higher density correspond to stiffer equations of state, which, in turn, enhance resistance to gravitational collapse. The compactness $u = M_{\text{eff}}/R$ and the associated surface redshift $z_s$ also exhibit smooth and physically acceptable behavior, remaining within the allowed relativistic limits.

The analysis of the energy conditions (NEC, WEC, SEC, DEC, and TEC) confirms that all are satisfied, validating the energy distribution, and ensuring that the matter content is physically realistic. The influence of the $f(Q)$ and EEH coupling parameters introduces a strong correlation between geometry and matter, enhancing the stiffness of the equation of state and allowing for more massive and stable compact configurations compared to standard general relativity. 

We also calculated the curves showing the Mass-Radius relation for different values of $\gamma$. Our analysis shows that for fixed mass, density, and other parameters, decreasing values of $\gamma$ show an increasing trend in the predicted radius of the star candidates.  

Overall, the results demonstrate a clear interdependence among the key stellar quantities; density, pressures, anisotropy, sound speeds, adiabatic index, and TOV forces, all of which collectively determine the equilibrium and stability of the system. When applied to the pulsar PSR~J1614--2230, the predicted mass and radius are found to be consistent with the observational data, confirming the astrophysical viability of the model. Therefore, the $f(Q)$-EEH gravity framework combined with the MIT Bag model provides a physically robust and observationally consistent description of strange quark stars, establishing a promising direction for future investigations of ultra-dense matter in modified gravity theories.

\end{document}